\documentclass[9pt,conference]{IEEEtran}
\usepackage{amssymb,amsthm,amsmath,array}
\usepackage{graphicx}
\usepackage[caption=false,font=footnotesize]{subfig}
\usepackage{xspace}
\usepackage[sort&compress, numbers]{natbib}
\usepackage{stmaryrd}
\usepackage{xcolor}
\usepackage{mathtools}
\usepackage{float}
\usepackage{textcomp}
\usepackage{amsmath}
\usepackage{amssymb,amsthm,amsmath,array}
\usepackage{graphicx}
\usepackage[caption=false,font=footnotesize]{subfig}
\usepackage{xspace}
\usepackage[sort&compress, numbers]{natbib}
\usepackage{stmaryrd}
\usepackage{xcolor}
\usepackage{mathtools}
\usepackage{float}
\usepackage{textcomp}

\usepackage{standalone}
\graphicspath{{Figures/}}

\usepackage{tikz}
\usetikzlibrary{external}
\usetikzlibrary{arrows}
\usepackage{pgfplots}
\pgfplotsset{compat=1.7}

\newcommand{\bs}{\boldsymbol}
\newcommand{\bb}{\mathbb}
\newcommand{\cl}{\mathcal}

\newcommand{\ts}{\textstyle}
\newcommand{\ie}{\emph{i.e.},\xspace}

\newcommand{\etal}{\emph{et al.}}
\newcommand{\sq}{\vspace{-2mm}}

\begin{document}
\title{Simultaneous High-Speed and Low-Dose 4-D STEM Using Compressive Sensing Techniques}
\author{\IEEEauthorblockN{
        Alex W. Robinson\IEEEauthorrefmark{1}\IEEEauthorrefmark{2},
        Amirafshar Moshtaghpour\IEEEauthorrefmark{1}\IEEEauthorrefmark{3},
        Jack Wells\IEEEauthorrefmark{2}\IEEEauthorrefmark{4},
        Daniel Nicholls\IEEEauthorrefmark{1}\IEEEauthorrefmark{2},\\
        Miaofang Chi\IEEEauthorrefmark{5},
        Ian MacLaren\IEEEauthorrefmark{6},
        Angus I. Kirkland\IEEEauthorrefmark{3}\IEEEauthorrefmark{7},
        Nigel D. Browning\IEEEauthorrefmark{1}\IEEEauthorrefmark{2}
    }
    \IEEEauthorblockA{
        \IEEEauthorrefmark{1} Department of Mechanical, Materials and Aerospace Engineering, University of Liverpool, UK.\\
        \IEEEauthorrefmark{2} SenseAI Innovations Ltd., University of Liverpool, Liverpool, UK.\\
        \IEEEauthorrefmark{3} Correlated Imaging Group, Rosalind Franklin Institute, Harwell Science and Innovation Campus, Didcot, UK.\\
        \IEEEauthorrefmark{4}Distributed Algorithms Centre for Doctoral Training, University of Liverpool, UK.\\
        \IEEEauthorrefmark{5}Chemical Science Division, Centre for Nanophase Materials Sciences, Oak Ridge National Laboratory, Tennessee, USA.\\
        \IEEEauthorrefmark{6}School of Physics and Astronomy, University of Glasgow, Glasgow, UK.\\
        \IEEEauthorrefmark{7} Department of Materials, University of Oxford, UK.}
}
\maketitle

\begin{abstract}
Here we show that compressive sensing allow 4-dimensional (4-D) STEM data to be obtained and accurately reconstructed with both high-speed and low fluence.  The methodology needed to achieve these results compared to conventional 4-D approaches requires only that a random subset of probe locations is acquired from the typical regular scanning grid, which immediately generates both higher speed and the lower fluence experimentally. We also consider downsampling of the detector, showing that oversampling is inherent within convergent beam electron diffraction (CBED) patterns, and that detector downsampling does not reduce precision but allows faster experimental data acquisition. Analysis of an experimental atomic resolution yttrium silicide data-set shows that it is possible to recover over 25dB peak signal-to-noise in the recovered phase using 0.3\% of the total data.  
\end{abstract}

\section{Introduction}\label{sec:intro}

\begin{figure*}[t!]
    \centering
    \scalebox{0.75}{\includegraphics{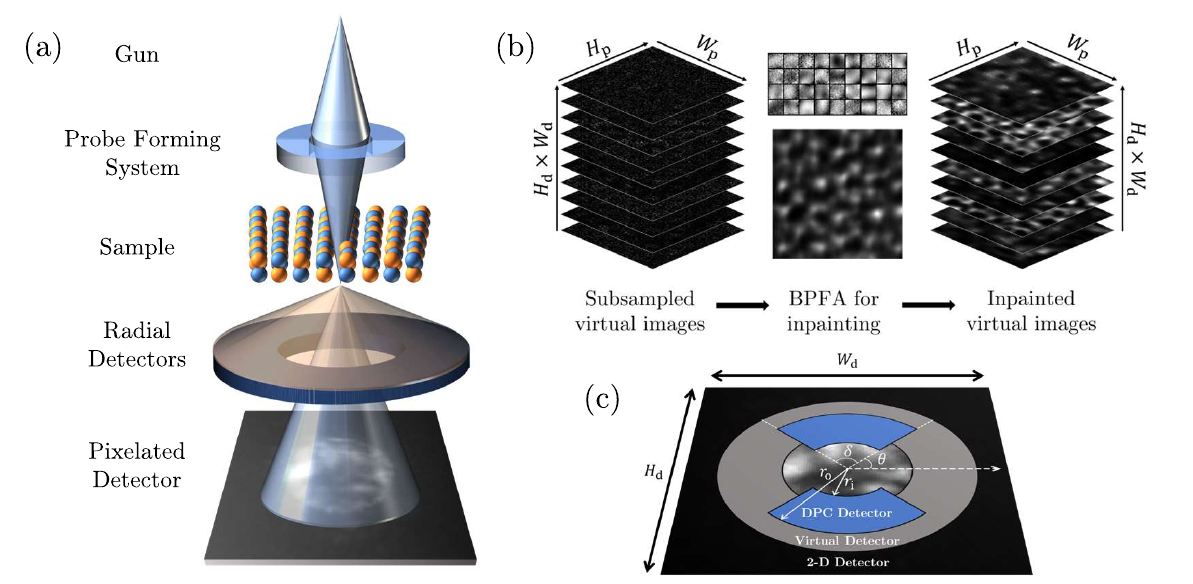}}
    \caption{Operating principles of 4-D STEM are demonstrated (a), electrons are converged to form a probe which is rastered in 2-D across the sample plane. The transmitted electrons are collected using a 2-D detector in the far field for each probe position. (b) Inpainting the 4-D STEM data-set by sequentially inpainting each virtual image using the BPFA algorithm. (c) Application of VDs and DPC at the detector plane.}\sq\sq\sq
    \label{fig:4-D STEM-scheme}
\end{figure*}

The goal of this study is to demonstrate that the application of a compressed acquisition methodology can improve the speed and reduce the fluence associated with 4-dimensional (4-D) scanning transmission electron microscopy (STEM). In this imaging mode a series of diffraction patterns for each probe position in a 2D grid are recorded in the far field on a 2D pixelated detector (Fig.~\ref{fig:4-D STEM-scheme})~\cite{ophus2019four}. Subsequently a variety of signals can be extracted by suitable geometric integration of regions at the detector.

Prior to the widespread use of aberration correctors, Nellist \etal demonstrated one of the earliest cases of 4-D STEM where coherent micro-diffraction patterns were collected as a function of probe position and used for a super-resolved ptychographic reconstruction~\cite{nellist1995resolution}. This allowed the resolution of the Si $\{004\}$ at $0.136$nm; a much higher spatial resolution than was achieveable using high-angle annular dark field (HAADF) STEM on the instrument used. Another early demonstration by Zaluzec \etal, used position resolved diffraction to image distributions of magnetic induction in a Lorentz STEM imaging mode~\cite{zaluzec2001lorentz,zaluzec2002quantitative}. 

4-D STEM has progressed significantly since these early demonstrations, with more recent examples of its application in ptychography having been used to recover the complex object wavefunction of weakly scattering objects, such as lithium ion cathode materials~\cite{lozano2018low} and biological samples~\cite{zhou2020low}. STEM ptychography has also been used to resolve praseodymium dumbbells at the limit set by thermal atomic motion~\cite{chen2021electron}. 4-D STEM has become popular due to its versatility by way of multi-modal imaging using virtual detectors (VDs)~\cite{ophus2019four}, differential phase contrast (DPC) ~\cite{shibata2012differential}, centre of mass (CoM) analysis~\cite{muller2017measurement}, and ptychography ~\cite{hoppe1969beugunga,hoppe1969beugungb,hegerl1970dynamische,hoppe1982trace,yang2016simultaneous}. A major limitation in the application of 4-D STEM has been the need for long integration times to a achieve significant signal-to-noise ratio (SNR) in the presence of noise and dark current. Although most commercially available direct electron detectors that operate in counting mode have effective frame rates of less than ten kHz, there have been recently announced direct electron detectors~\cite{faruqi2005direct, ryll2016pnccd,faruqi2018direct,ciston20194d} operating at between 100kHz and 1MHz, albeit with small pixel array sizes. Using these detectors CBED patterns can be acquired with little or no noise at an effective dwell time of 10µs per probe position~\cite{philipp2022very,ciston20194d}.  While these are significant improvements over earlier indirect scintillator coupled detectors operating at fewer than $30$ fps~\cite{maclaren2020detectors,faruqi2007electronic}, it remains the case that only the most recent detectors match the dwell time of traditional solid state monolithic STEM detectors. Importantly, our approach can also be used with slower large pixel array detectors to provide the required matching speed increase.

Hence, 4-D STEM experiments remain susceptible to drift and beam induced damage~\cite{egerton2021radiation} which potentially limits its applicability to studies of beam sensitive organic and hybrid materials or to investigations of materials dynamics.

One option to overcome beam damage is to reduce the electron fluence at the sample~\cite{bustillo20214d, li20224d}. By reducing the fluence below a materials dependent threshold~\cite{egerton2021dose}, or by using cryogenic temperatures~\cite{zhou2020low}, beam damage can be reduced. Furthermore, if combined with alternative methods to increase acquisition speeds such as low bit-depth electron counting~\cite{yang20154d,o2020phase}, the acquisition speed can be increased and sample drift can be reduced. However, given that the SNR is related to the number of detected electrons, and hence, with the fluence per probe position, a combination of fluence and fast acquisition quickly transitions the experiment to conditions that are below the minimum signal-to-noise requirements for 4-D methods such as ptychography~\cite{pennycook2019high}.  An alternative method to overcome beam damage (as well as to increase the effective frame rate of an existing detector) in STEM is by using techniques based on the theory of compressive sensing (CS)~\cite{donoho2006compressed,candes2006robust}, which is referred to here as probe sub-sampling. Probe sub-sampling in this context refers to controlling the set of positions of the STEM probe visits within a raster scan to reduce the number of acquisition points- thereby directly creating a faster scan and a lower fluence and flux at the sample. Probe sub-sampling has already been experimentally demonstrated for a variety of experimental STEM and SEM imaging modes~\cite{binev2012compressed,anderson2013sparse,stevens2014potential,stevens2018subsampled,nicholls2020minimising,mehdi2019controlling,nicholls2021subsampled,nicholls2022compressive,nicholls2022targeted}, and has also been used to speed up the computational time for STEM simulations~\cite{robinson2022sim,robinson2022compressed,robinson2023towards}. The key benefit for probe sub-sampling in STEM is that by acquiring less data, acquisition rates can be increased, which in turn reduces drift artefacts as well as reducing the total cumulative electron fluence of the entire field of view. Thus, samples which are susceptible to beam damage can be imaged at usable SNRs, without over exposure to the incident beam. Although the dose at any acquired probe location is independent of the scan pattern, work by Nicholls \etal~\cite{nicholls2020minimising} has shown that the diffusion of radicals due to beam interactions at neighbouring probe locations compounds the damage of samples. By taking larger steps in a random fashion, this cumulative dose can be reduced since radicals are not propagated between successive probe locations.
In this paper, we will demonstrate a focused probe acquisition method which reduces beam damage and increases acquisition rate by probe sub-sampling. We acquire only a subset of the CBED patterns and use a Bayesian dictionary learning technique known as Beta Process Factor Analysis (BPFA) to recover the full 4-D STEM data-set from the sub-sampled measurements. The BPFA has been shown as a robust inpainting algorithm to data containing complex structures such as defects~\cite{robinson2023towards}, and further evidence is given in the Supplemental Material. We describe simulations of this method to a 4-D STEM data-set of yttrium silicide, and demonstrate that 4-D STEM data acquisition can be reduced by at least $256\times$ without significant quality loss in all imaging modes. 

Previous work by Stevens \etal~\cite{stevens2018subsampled}, demonstrated that with probe sub-sampling and detector sub-sampling can be employed and that by inpainting followed by phase retrieval, one can recover functionally identical~\footnote{Functionally identical results are defined as the preservation of features compared to the ground truth, such that the analysis is preserved in determining properties of the sample.} results to a fully sampled experiment. In this work the inpainting of the 4-D data used a Kruskal-factor analysis technique~\cite{stevens2017tensor}. We extend this approach by using a new implementation of the BPFA algorithm which takes advantage of GPU acceleration. We will also build on the work of Zhang \etal ~\cite{zhang2021many} who showed that the number of detector pixels required for ptychographic reconstruction can be reduced significantly without loss of resolution.

\section{Proposed method for sub-sampled 4-D STEM}
\begin{figure*}[t!]
    \centering
    \scalebox{0.8}{\includegraphics{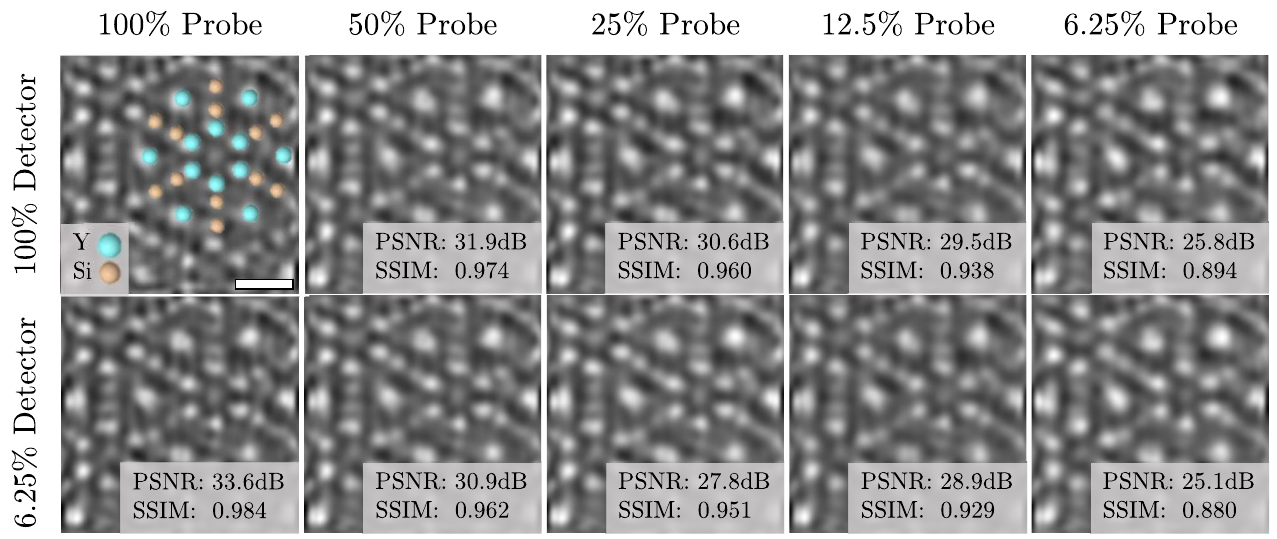}}
    \caption{Visual comparison of ptychographic phase retrieval quality for different probe sub-sampling and detector downsampling ratios. The reference data is the full data-set passed through the BPFA algorithm (top row, leftmost column). The scale bar indicates 0.5nm.}\sq\sq
    \label{fig:y5si3}
\end{figure*}
The experimental set-up for the acquisition of a sub-sampled data-set is shown in Fig.~\ref{fig:4-D STEM-scheme}. We assume a pixelated detector with $H_{\rm d}$ and $W_{\rm d}$ pixels in the vertical and horizontal axis, respectively, collecting 2-D CBED patterns of size $H_{\rm d} \times W_{\rm d}$. Let $\Omega_{\rm d} \coloneqq \{1,\cdots, H_{\rm d}\}\times \{1,\cdots, W_{\rm d}\}$ be the set of all detector pixel locations and $\bs k_{\rm d} \coloneqq (k^{\rm h}_{\rm d}, k^{\rm w}_{\rm d}) \in \Omega_{\rm d}$ denote the coordinates of a detector pixel. We further assume an electron probe scanning a regular grid of $H_{\rm p}$ and $W_{\rm p}$ locations in the vertical and horizontal axis, respectively~\footnote{Note that the coordinate axes of the pixelated detector and scanning probe are not necessarily the same.}, collected in a probe locations set $\Omega_{\rm p} \coloneqq \{1,\cdots,H_{\rm p}\} \times \{1,\cdots,W_{\rm p}\}$. Let $\bs r_{\rm p} \coloneqq (r^{\rm h}_{\rm p}, r^{\rm w}_{\rm p}) \in \Omega_{\rm p}$ denote the coordinates of a probe location. Moreover, the total number of detector pixels and probe locations are denoted by, respectively, $N_{\rm p} = H_{\rm p} W_{\rm p}$ and $N_{\rm d} = H_{\rm d} W_{\rm d}$. Finally, given a scan step parameter $\Delta_{\rm p}$, in m, of the electron probe and detector pixel size $\Delta_{\rm d}$, in mrad, the location of the scanning probe and detector pixel can be converted from their index units to real units.

Let $\cl X\in \bb R^{H_{\rm p}\times W_{\rm p}\times H_{\rm d} \times W_{\rm d}}$ be the discretised 4-D representation of fully sampled 4-D STEM data; and $\cl X(\bs r_{\rm p}, \bs k_{\rm d})$ be the 4-D STEM data observed at probe location $\bs r_{\rm p}$ and detector pixel $\bs k_{\rm d}$. A \textit{CBED pattern} collected at probe location $\bs r_{\rm p}$ is denoted by $\bs X^{\rm dp}_{\bs r_{\rm p}} \coloneqq \cl X(\bs r_{\rm p}, \cdot) \in \bb R^{H_{\rm d} \times W_{\rm d}}$. In this paper, the \textit{virtual image} corresponding to a detector pixel $\bs k_{\rm d}$, represented as $\bs X^{\rm vi}_{\bs k_{\rm d}} \coloneqq \cl X(\cdot, \bs k_{\rm d}) \in \bb R^{H_{\rm p} \times W_{\rm p}}$, refers to a matrix collecting the data observed at detector pixel $\bs k_{\rm d}$ for all probe positions.

We achieve our compressed 4-D STEM by sub-sampling $M_{\rm p} \ll N_{\rm p}$ probe locations acquired in the sub-sampling set $\Omega\subset\Omega_{\rm p}$, which is equivalent to sub-sampling each of the virtual images (sharing a common mask determined by $\Omega$). This defines our acquisition model as,
\begin{equation}\label{eq:cseq}
{\bs Y}^{\rm vi}_{\bs k_{\rm d}} = {\bs P}_{\Omega} ({\bs X}^{\rm vi}_{\bs k_{\rm d}}) + {\bs N}_{\bs k_{\rm d}} \in \bb R^{H_{\rm p} \times W_{\rm p}}, \quad {\rm for~} \bs k_{\rm d} \in \Omega_{\rm d},
\end{equation} 
where ${\bs Y}^{\rm vi}_{\bs k_{\rm d}}$ is the sub-sampled measurements at detector pixel ${\bs k_{\rm d}}$ and ${\bs P}_{\Omega}$ is a mask operator with $({\bs P}_{\Omega} (\bs U))_{(i,j)} = {\bs U}_{(i,j)}$ if $(i,j)\in\Omega$ and $({\bs P}_{\Omega} (\bs U))_{(i,j)} = 0$ otherwise, and ${\bs N}_{\bs k_{\rm d}}$ is an additive noise.

We now estimate virtual images $\hat{\bs X}^{\rm vi}_{\bs k_{\rm d}} \approx {\bs X}^{\rm vi}_{\bs k_{\rm d}}$ from sub-sampled measurements ${\bs Y}^{\rm vi}_{\bs k_{\rm d}}$ in \eqref{eq:cseq} for $\bs k_{\rm d} \in \Omega_{\rm d}$, which defines the inpainting problem. In this work we assume that virtual images are sparse or compressible\footnote{A signal is sparse if it strictly contains only a few non-zero weights in a dictionary, whereas a signal is compressible if the magnitudes of the weights decay rapidly when in descending order.} in an unknown dictionary that can be learned during the recovery process. This leads to the development of dictionary learning adopting a Bayesian non-parametric method called Beta-Process Factor Analysis (BPFA) as introduced in ~\cite{paisley2009nonparametric}. The advantages of this approach include the ability to infer both the noise variance and sparsity level of the signal in the dictionary, and allows for the learning of dictionary elements directly from sub-sampled data. This approach has been tested in previous reports~\cite{nicholls2021subsampled,nicholls2022compressive,robinson2022sim,nicholls2022targeted,robinson2023towards} and has shown success when applied to electron microscopy data. Note that this approach learns a different dictionary for each virtual image and a BPFA instance is applied to every virtual image. This is not necessarily optimal, however, we will leave the concept of learning a shared dictionary for all virtual images and applying a single instance of BPFA directly on the sub-sampled 4-D data to a future study (a full description of the BPFA process is provided in the Supplemental Material\footnote{See Supplemental Material at [URL will be inserted by publisher] for a full description of the BPFA process}).

\begin{figure}[b!]
           \centering
            \scalebox{0.55}{\includegraphics{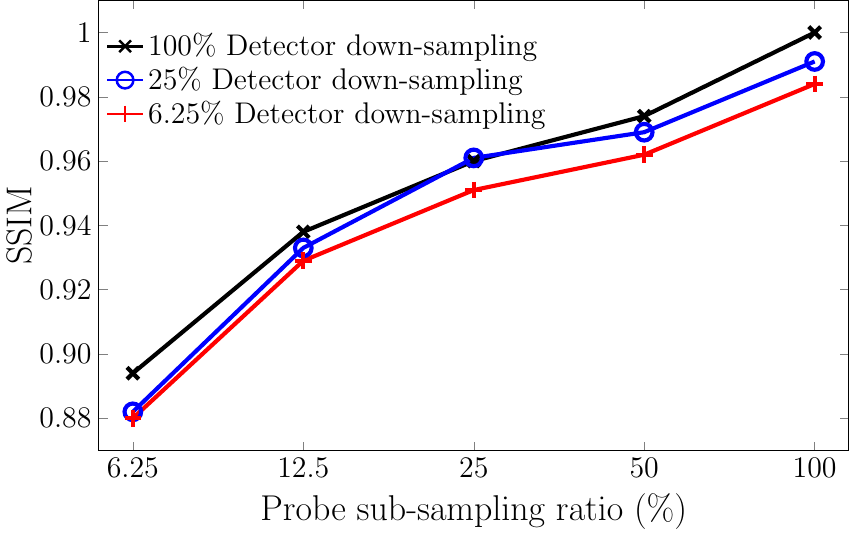}}
           \caption{SSIM of phases with respect to probe and detector sampling ratios. As the probe sub-sampling ratio increases, the quality of the phase increases. However, there is only a small difference in the phase quality as the detector downsampling ratio is decreased. This indicates significant redundancy within the 4-D data-set, which can be omitted through detector downsampling and probe sub-sampling. Example images of this experiment are shown in Fig.~\ref{fig:y5si3}.}\sq\sq
           \label{fig:phaseplot}
\end{figure}

In addition to probe sub-sampling, we can also downsample the detector pixels to eliminate redundancy. This can also be inferred as the optimisation of our reciprocal space sampling, $\Delta_{\rm d}$, which can be carried out by only reading out the set of rows which are within the sampling set. This is different to conventional detector pixel binning (which still requires reading of all rows within the total CBED pattern), since we do not consider nor acquire rows which do not belong to the sampling set.

\begin{figure*}[t!]
    \centering
\begin{minipage}{\textwidth}
      \centering
      \begin{minipage}{0.42\linewidth}
      
          \begin{figure}[H]
             \centering
             \scalebox{0.5}{\includegraphics{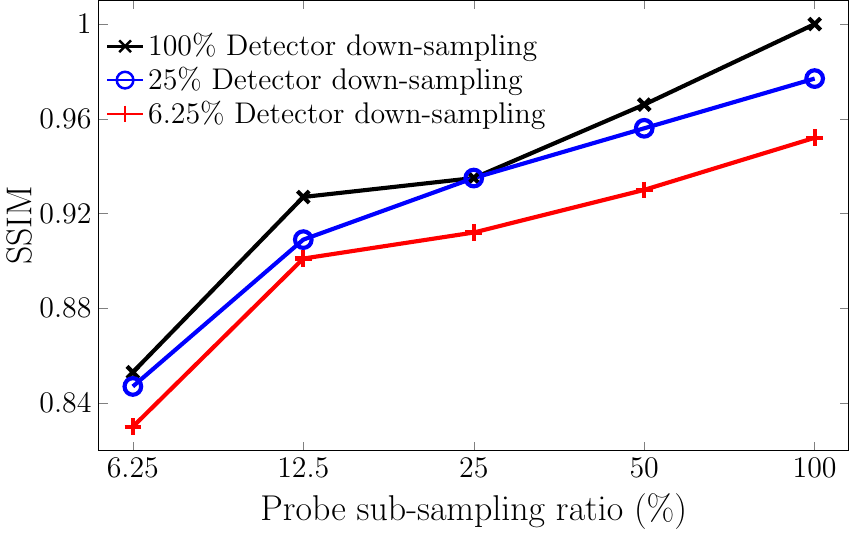}}\\
             (a) SSIM of CoM fields with respect to probe and detector sampling ratios.
            
        \end{figure}
      \end{minipage}
      \hspace{0.0001\linewidth}
      \begin{minipage}{0.55\linewidth}
          \begin{figure}[H]
            \centering
            \scalebox{0.54}{\includegraphics{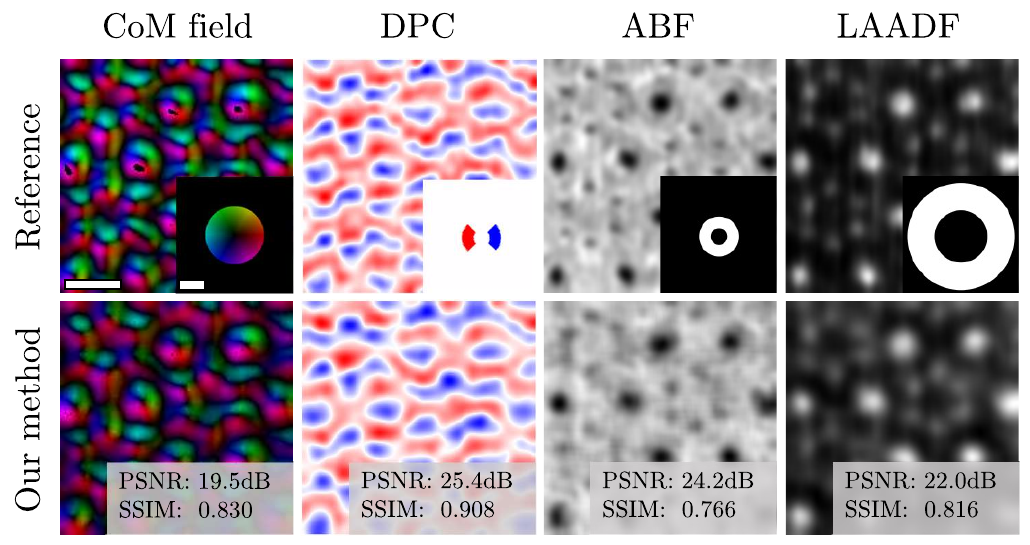}}\\
            (b) VDs (ABF and LAADF), DPC, and CoM analysis. The DPC detector subtracts the integral of the blue from the red region.
            \end{figure}
      \end{minipage}
  \end{minipage}
  \caption{(a) SSIM values as a quality metric for CoM field images. (b) CoM field, DPC, ABF, and LAADF images for $6.25\%$ probe sampling and $6.25\%$ detector downsampling after inpainting. The reference data is the full data passed through the BPFA algorithm (top row). The PSNR and SSIM values are overlaid, the spatial scale bar indicates 0.5nm, and the detector scale bar indicates $30$ mrad. The left-most column is an example data-point from the plot in (a), and the corresponding plots similar to (a) for DPC, ABF, and LAADF can be found in the Supplemental Material.}\sq\sq\label{fig:comfieldplot}
 \end{figure*}
Given the detector downsampling factor $f_{\rm d} \in \bb N$, we first uniformly read-out every $f_{\rm d}^{\rm th}$ row on the detector. This results in faster acquisition of CBED patterns of size $H_{\rm d}/f_{\rm d} \times W_{\rm d}$ pixels. To further reduce the size of the data-set, we then keep only the data from every $f_{\rm d}^{\rm th}$ column on the detector; resulting in CBED patterns with $M_{\rm d} = H_{\rm d}\cdot W_{\rm d}/f^2_{\rm d}$ entries. In this paper, we define detector downsampling ratio as $M_{\rm d}/N_{\rm d} = 1/f^2_{\rm d}$. In practice, the camera length could also be varied to optimise $\Delta_{\rm{d}}$ since the camera length is inversely proportional to the reciprocal space sampling. This would account for detectors which cannot read out rows/pixels independently. It would also effectively bin the signal on the detector where hardware binning is limited, improving signal-to-noise.

\section{Results}

In order to model experimental acquisition, an experimental 4-D STEM data-set of Y$_{5}$Si$_{3}$ was used (with all scan positions) and applied random sub-sampling of the probe positions and downsampling of the CBED patterns. 

Y$_{5}$Si$_{3}$ is an electride framework composed of cation and anion sublattices. These sublattices have a net positive electric charge which are balanced by loosely bonded, interstitial anionic electrons~\cite{zheng2021direct}. Y$_{5}$Si$_{3}$ has been proposed as a low Schottky barrier material for \textit{n}-type silicon semiconductors due to its low Schottky barrier height of $0.27$eV~\cite{isogai2008formation}. Readers are referred to Zheng \etal~\cite{zheng2021direct} for details on practical applications.  

The experimental data was acquired using a probe forming aperture semi-angle of 30mrad from a 100kV electron electron source with a probe current of $20$pA with a dwell-time of 1.3ms. A $\Delta_{\rm{p}}$ of $0.0108$nm~was used, giving a theoretical electron fluence of approximately $1.4\times 10^{9}$e$^{-}$nm$^{-2}$. The camera collected diffraction patterns of size $128\times 128$ pixels, where $\Delta_{\rm{d}}$ is $1$mrad. In this study we applied probe sub-sampling ratios $M_{\rm p}/N_{\rm p} \in \{6.25, 12.5, 25, 50, 100\}\%$, as well as detector downsampling ratios $M_{\rm d}/N_{\rm d} \in\{6.25, 25, 100\}\%$. LAADF and annular BF (ABF)~\cite{okunishi2012experimental} virtual detector images, $(r_{\rm i},~r_{\rm o}) = (30,~60)$ mrad and $(r_{\rm i},~r_{\rm o}) = (10,~22)$ mrad were simulated together with DPC images with $(r_{\rm i},~r_{\rm o}) = (10,~22)$ mrad and $(\theta,~\delta) = (3\pi/4,~\pi/2)$ rad.  In addition we simulated the recovered ptychographic phase (Fig.~\ref{fig:y5si3}). For this there are a number of analytical and iterative algorithms~\cite{rodenburg2004phase,maiden2009improved,maiden2012annealing,maiden2012ptychographic,elser2003phase,d2014deterministic} that recover the complex ptychographic wavefunction, and here we used a modification of the Wigner distribution deconvolution (WDD) algorithm~\cite{bates1989sub,rodenburg1992theory,yang2017electron,martinez2017comparison,lozano2018low,o2021contrast} within the \textit{ptychoSTEM} package for MATLAB~\cite{yang2016simultaneous}. Details on the analysis methods used can be found in the Supplemental Material.

Fig.~\ref{fig:phaseplot} shows the quality of the ptychographic phase (using the structural similarity index measure (SSIM)~\cite{wang2004image} as our chosen metric) with respect to different probe sub-sampling and detector downsampling ratios. There is only a small degradation in the quality as the sampling at the detector is decreased; this implies the detector is over-sampled. We further observe that probe sub-sampling can be used with BPFA to recover visually identical results in the phase. 

Similarly, Fig.~\ref{fig:comfieldplot}(a) shows a comparison of the quality of CoM field analysis as a function of sub-sampling ratio, where visually identical results are achieved with respect to the reference data. Comparing Fig.~\ref{fig:phaseplot} and Fig.~\ref{fig:comfieldplot}(a) suggests that ptychographic phase recovery is more robust in this case. This is possibly due to the fact that the WDD operates on a full 4-D data-set, while the CoM field is computed from individual CBED patterns.

Fig.~\ref{fig:comfieldplot}(b) is a direct image comparison between our reference data and reduced sampling data ($M_{\rm p}/N_{\rm p} = M_{\rm d}/N_{\rm d} = 6.25\%$) when applied to CoM field analysis, DPC, ABF, and LAADF.  It is clear that there is very little difference in the quality of the images from a visual perspective, and this is supported comparison of the corresponding peak signal-to-noise (PSNR) and SSIM values corresponding to each. Fig.~\ref{fig:y5si3} is a visual comparison of the data in Fig.~\ref{fig:phaseplot}. As can be seen, the recovered phase data is almost indistinguishable, with all showing the expected location of yttrium and silicon atoms. 

\section{Conclusions}

Our results demonstrate the inherent redundancy within the 4-D STEM data-set. By utilising inpainting algorithms, it is possible to discard over $99.6\%$ (see Fig.~\ref{fig:y5si3} bottom-right) of the original data-set whilst still recovering qualitatively identical results in the reconstructed phase, CoM field, DPC and VD images, to those obtained from processing the full data-set. 

This method has also been shown as robust to 4-D STEM data containing an interface, and the results are given in Fig.~S5 in the Supplemental Material.

However, given the inherent redundancy in 4-D STEM data, we propose that even lower sampling ratios could be employed using a multi-dimensional recovery algorithm. The benefit of this is that by using a multi-dimensional recovery algorithm we can leverage more data during the training process as well as the similarity between virtual images during the recovery step. It may be possible to also include sparse detector sampling followed by inpainting the 4-D STEM data-set with minor modifications to the acquisition model. This could further increase acquisition speeds by assuming that each pixel has a fixed read-out time, and potentially allow for multiple 4-D STEM data-sets to be acquired rapidly. We postulate that time-resolved 4-D STEM is now not limited by the detector read-out speed, but can instead be acquired through reduced sampling strategies.
\section{Acknowledgments}
 This work was performed at the Albert Crewe Centre (ACC) for Electron Microscopy, a shared research facility (SRF) fully supported by the University of Liverpool. This work was also funded by the EPSRC Centre for Doctoral Training in Distributed Algorithms (EP/S023445/1), Sivananthan Labs, and the Rosalind Franklin Institute. M.C. would like to acknowledge the support by the US DOE Office of Science Early Career project FWP\# ERKCZ55 and the Center for Nanophase Materials Sciences (CNMS), a US DOE Office of Science User Facility. Initial experiments were carried out using MagTEM, a JEOL ARM200F STEM in the Kelvin Nanocharacterisation Centre , which was installed with support from the University of Glasgow and the Scottish Universities Physics Alliance.  A.W.R. would like to thank Jordan A. Hatchel (ORNL) for his knowledge and insights of 4-D STEM analysis.

\section*{Data Availability Statement}
The data that support the findings of this study are available within the article and its supplementary material.

\bibliographystyle{IEEEtran}
\bibliography{refs}

\begin{thebibliography}{10}
\providecommand{\url}[1]{#1}
\csname url@samestyle\endcsname
\providecommand{\newblock}{\relax}
\providecommand{\bibinfo}[2]{#2}
\providecommand{\BIBentrySTDinterwordspacing}{\spaceskip=0pt\relax}
\providecommand{\BIBentryALTinterwordstretchfactor}{4}
\providecommand{\BIBentryALTinterwordspacing}{\spaceskip=\fontdimen2\font plus
\BIBentryALTinterwordstretchfactor\fontdimen3\font minus \fontdimen4\font\relax}
\providecommand{\BIBforeignlanguage}[2]{{%
\expandafter\ifx\csname l@#1\endcsname\relax
\typeout{** WARNING: IEEEtran.bst: No hyphenation pattern has been}%
\typeout{** loaded for the language `#1'. Using the pattern for}%
\typeout{** the default language instead.}%
\else
\language=\csname l@#1\endcsname
\fi
#2}}
\providecommand{\BIBdecl}{\relax}
\BIBdecl

\bibitem{ophus2019four}
C.~Ophus, ``Four-dimensional scanning transmission electron microscopy ({4D-STEM}): {F}rom scanning nanodiffraction to ptychography and beyond,'' \emph{Microscopy and Microanalysis}, vol.~25, no.~3, pp. 563--582, 2019.

\bibitem{nellist1995resolution}
P.~Nellist, B.~McCallum, and J.~M. Rodenburg, ``Resolution beyond the information limit in transmission electron microscopy,'' \emph{Nature}, vol. 374, no. 6523, pp. 630--632, 1995.

\bibitem{zaluzec2001lorentz}
N.~J. Zaluzec, ``Lorentz {STEM}: A digital approach to an old technique,'' \emph{Microscopy and Microanalysis}, vol.~7, no.~S2, pp. 222--223, 2001.

\bibitem{zaluzec2002quantitative}
------, ``Quantitative measurements of magnetic vortices using position resolved diffraction in {L}orentz {STEM},'' \emph{Microscopy and Microanalysis}, vol.~8, no. S02, pp. 376--377, 2002.

\bibitem{lozano2018low}
J.~G. Lozano, G.~T. Martinez, L.~Jin, P.~D. Nellist, and P.~G. Bruce, ``Low-dose aberration-free imaging of {L}i-rich cathode materials at various states of charge using electron ptychography,'' \emph{Nano letters}, vol.~18, no.~11, pp. 6850--6855, 2018.

\bibitem{zhou2020low}
L.~Zhou, J.~Song, J.~Kim, X.~Pei, C.~Huang, M.~Boyce, L.~Mendonça, D.~Clare, A.~Siebert, C.~Allen, E.~Liberti, D.~Stuart, X.~Pan, P.~Nellist, P.~Zhang, A.~Kirkland, and P.~Wang, ``Low-dose phase retrieval of biological specimens using cryo-electron ptychography,'' \emph{Nature communications}, vol.~11, no.~1, pp. 1--9, 2020.

\bibitem{chen2021electron}
Z.~Chen, Y.~Jiang, Y.-T. Shao, M.~E. Holtz, M.~Odstr{\v{c}}il, M.~Guizar-Sicairos, I.~Hanke, S.~Ganschow, D.~G. Schlom, and D.~A. Muller, ``Electron ptychography achieves atomic-resolution limits set by lattice vibrations,'' \emph{Science}, vol. 372, no. 6544, pp. 826--831, 2021.

\bibitem{shibata2012differential}
N.~Shibata, S.~D. Findlay, Y.~Kohno, H.~Sawada, Y.~Kondo, and Y.~Ikuhara, ``Differential phase-contrast microscopy at atomic resolution,'' \emph{Nature Physics}, vol.~8, no.~8, pp. 611--615, 2012.

\bibitem{muller2017measurement}
K.~Müller-Caspary, F.~F. Krause, T.~Grieb, S.~Löffler, M.~Schowalter, A.~Béché, V.~Galioit, D.~Marquardt, J.~Zweck, P.~Schattschneider, J.~Verbeeck, and A.~Rosenauer, ``Measurement of atomic electric fields and charge densities from average momentum transfers using scanning transmission electron microscopy,'' \emph{Ultramicroscopy}, vol. 178, pp. 62--80, 2017.

\bibitem{hoppe1969beugunga}
W.~Hoppe, ``Beugung im inhomogenen {P}rim{\"a}rstrahlwellenfeld. {III}. {A}mplituden-und {P}hasenbestimmung bei unperiodischen {O}bjekten,'' \emph{Acta Crystallographica Section A: Crystal Physics, Diffraction, Theoretical and General Crystallography}, vol.~25, no.~4, pp. 508--514, 1969.

\bibitem{hoppe1969beugungb}
W.~Hoppe and G.~Strube, ``Beugung in inhomogenen {P}rim{\"a}rstrahlenwellenfeld. {II}. {L}ichtoptische analogieversuche zur {P}hasenmessung von gitterinterferenzen,'' \emph{Acta Crystallographica Section A: Crystal Physics, Diffraction, Theoretical and General Crystallography}, vol.~25, no.~4, pp. 502--507, 1969.

\bibitem{hegerl1970dynamische}
R.~Hegerl and W.~Hoppe, ``Dynamische theorie der kristallstrukturanalyse durch elektronenbeugung im inhomogenen prim{\"a}rstrahlwellenfeld,'' \emph{Berichte der Bunsengesellschaft f{\"u}r physikalische Chemie}, vol.~74, no.~11, pp. 1148--1154, 1970.

\bibitem{hoppe1982trace}
W.~Hoppe, ``Trace structure analysis, ptychography, phase tomography,'' \emph{Ultramicroscopy}, vol.~10, no.~3, pp. 187--198, 1982.

\bibitem{yang2016simultaneous}
H.~Yang, R.~Rutte, L.~Jones, M.~Simson, R.~Sagawa, H.~Ryll, M.~Huth, T.~Pennycook, M.~Green, H.~Soltau, Y.~Kondo, B.~Davis, and P.~Nellist, ``Simultaneous atomic-resolution electron ptychography and {Z}-contrast imaging of light and heavy elements in complex nanostructures,'' \emph{Nature Communications}, vol.~7, no.~1, pp. 1--8, 2016.

\bibitem{faruqi2005direct}
A.~Faruqi, R.~Henderson, M.~Pryddetch, P.~Allport, and A.~Evans, ``Direct single electron detection with a {CMOS} detector for electron microscopy,'' \emph{Nuclear Instruments and Methods in Physics Research Section A: Accelerators, Spectrometers, Detectors and Associated Equipment}, vol. 546, no. 1-2, pp. 170--175, 2005.

\bibitem{ryll2016pnccd}
H.~Ryll, M.~Simson, R.~Hartmann, P.~Holl, M.~Huth, S.~Ihle, Y.~Kondo, P.~Kotula, A.~Liebel, K.~Müller-Caspary, A.~Rosenauer, R.~Sagawa, J.~Schmidt, H.~Soltau, and L.~Strüder, ``A pn{CCD}-based, fast direct single electron imaging camera for {TEM} and {STEM},'' \emph{Journal of Instrumentation}, vol.~11, no.~04, p. P04006, 2016.

\bibitem{faruqi2018direct}
A.~Faruqi and G.~McMullan, ``Direct imaging detectors for electron microscopy,'' \emph{Nuclear Instruments and Methods in Physics Research Section A: Accelerators, Spectrometers, Detectors and Associated Equipment}, vol. 878, pp. 180--190, 2018.

\bibitem{ciston20194d}
J.~Ciston, I.~J. Johnson, B.~R. Draney, P.~Ercius, E.~Fong, A.~Goldschmidt, J.~M. Joseph, J.~R. Lee, A.~Mueller, C.~Ophus, A.~Selvarajan, D.~E. Skinner, T.~Stezelberger, C.~S. Tindall, A.~M. Minor, and P.~Denes, ``The {4D} camera: {V}ery high speed electron counting for {4D-STEM},'' \emph{Microscopy and Microanalysis}, vol.~25, no.~S2, pp. 1930--1931, 2019.

\bibitem{philipp2022very}
H.~T. Philipp, M.~W. Tate, K.~S. Shanks, L.~Mele, M.~Peemen, P.~Dona, R.~Hartong, G.~van Veen, Y.-T. Shao, Z.~Chen, J.~Thom-Levy, D.~A. Muller, and S.~M. Gruner, ``Very-high dynamic range, 10,000 frames/second pixel array detector for electron microscopy,'' \emph{Microscopy and Microanalysis}, vol.~28, no.~2, pp. 425--440, 2022.

\bibitem{maclaren2020detectors}
I.~MacLaren, T.~A. Macgregor, C.~S. Allen, and A.~I. Kirkland, ``Detectors— the ongoing revolution in scanning transmission electron microscopy and why this important to material characterization,'' \emph{{APL} {M}aterials}, vol.~8, no.~11, p. 110901, 2020.

\bibitem{faruqi2007electronic}
A.~Faruqi and R.~Henderson, ``Electronic detectors for electron microscopy,'' \emph{Current opinion in structural biology}, vol.~17, no.~5, pp. 549--555, 2007.

\bibitem{egerton2021radiation}
R.~Egerton, ``{R}adiation {D}amage and {N}anofabrication in {TEM} and {STEM},'' \emph{Microscopy Today}, vol.~29, no.~3, p. 56–59, 2021.

\bibitem{bustillo20214d}
K.~C. Bustillo, S.~E. Zeltmann, M.~Chen, J.~Donohue, J.~Ciston, C.~Ophus, and A.~M. Minor, ``{4D-STEM} of beam-sensitive materials,'' \emph{Accounts of Chemical Research}, vol.~54, no.~11, pp. 2543--2551, 2021.

\bibitem{li20224d}
G.~Li, H.~Zhang, and Y.~Han, ``{4D-STEM} {P}tychography for {E}lectron-{B}eam-{S}ensitive {M}aterials,'' \emph{{ACS} Central Science}, 2022.

\bibitem{egerton2021dose}
R.~Egerton, ``Dose measurement in the {TEM} and {STEM},'' \emph{Ultramicroscopy}, vol. 229, p. 113363, 2021.

\bibitem{yang20154d}
H.~Yang, L.~Jones, H.~Ryll, M.~Simson, H.~Soltau, Y.~Kondo, R.~Sagawa, H.~Banba, I.~MacLaren, and P.~Nellist, ``{4D STEM}: {H}igh efficiency phase contrast imaging using a fast pixelated detector,'' in \emph{Journal of Physics: Conference Series}, vol. 644, no.~1.\hskip 1em plus 0.5em minus 0.4em\relax IOP Publishing, 2015, p. 012032.

\bibitem{o2020phase}
C.~O'Leary, C.~Allen, C.~Huang, J.~Kim, E.~Liberti, P.~Nellist, and A.~Kirkland, ``Phase reconstruction using fast binary {4D STEM} data,'' \emph{Applied Physics Letters}, vol. 116, no.~12, p. 124101, 2020.

\bibitem{pennycook2019high}
T.~J. Pennycook, G.~T. Martinez, P.~D. Nellist, and J.~C. Meyer, ``High dose efficiency atomic resolution imaging via electron ptychography,'' \emph{Ultramicroscopy}, vol. 196, pp. 131--135, 2019.

\bibitem{donoho2006compressed}
D.~L. Donoho, ``Compressed sensing,'' \emph{IEEE transactions on information theory}, vol.~52, no.~4, pp. 1289--1306, 2006.

\bibitem{candes2006robust}
E.~J. Cand{\`e}s, J.~Romberg, and T.~Tao, ``Robust uncertainty principles: Exact signal reconstruction from highly incomplete frequency information,'' \emph{IEEE transactions on information theory}, vol.~52, no.~2, pp. 489--509, 2006.

\bibitem{binev2012compressed}
P.~Binev, W.~Dahmen, R.~DeVore, P.~Lamby, D.~Savu, and R.~Sharpley, ``Compressed sensing and electron microscopy,'' in \emph{Modelling Nanoscale Imaging in Electron Microscopy}.\hskip 1em plus 0.5em minus 0.4em\relax Springer, 2012, pp. 73--126.

\bibitem{anderson2013sparse}
H.~S. Anderson, J.~Ilic-Helms, B.~Rohrer, J.~Wheeler, and K.~Larson, ``Sparse imaging for fast electron microscopy,'' in \emph{Computational Imaging XI}, vol. 8657.\hskip 1em plus 0.5em minus 0.4em\relax SPIE, 2013, pp. 94--105.

\bibitem{stevens2014potential}
A.~Stevens, H.~Yang, L.~Carin, I.~Arslan, and N.~D. Browning, ``The potential for {B}ayesian compressive sensing to significantly reduce electron dose in high-resolution {STEM} images,'' \emph{Microscopy}, vol.~63, no.~1, pp. 41--51, 2014.

\bibitem{stevens2018subsampled}
A.~Stevens, H.~Yang, W.~Hao, L.~Jones, C.~Ophus, P.~D. Nellist, and N.~D. Browning, ``Subsampled {STEM}-ptychography,'' \emph{Applied Physics Letters}, vol. 113, no.~3, p. 033104, 2018.

\bibitem{nicholls2020minimising}
D.~Nicholls, J.~Lee, H.~Amari, A.~J. Stevens, B.~L. Mehdi, and N.~D. Browning, ``Minimising damage in high resolution scanning transmission electron microscope images of nanoscale structures and processes,'' \emph{Nanoscale}, vol.~12, no.~41, pp. 21\,248--21\,254, 2020.

\bibitem{mehdi2019controlling}
B.~L. Mehdi, A.~Stevens, L.~Kovarik, N.~Jiang, H.~Mehta, A.~Liyu, S.~Reehl, B.~Stanfill, L.~Luzi, W.~Hao, L.~Bramer, and N.~D. Browning, ``Controlling the spatio-temporal dose distribution during {STEM} imaging by subsampled acquisition: {I}n-situ observations of kinetic processes in liquids,'' \emph{Applied Physics Letters}, vol. 115, no.~6, p. 063102, 2019.

\bibitem{nicholls2021subsampled}
D.~Nicholls, J.~Wells, A.~Stevens, Y.~Zheng, J.~Castagna, and N.~D. Browning, ``Sub-sampled imaging for {STEM}: Maximising image speed, resolution and precision through reconstruction parameter refinement,'' \emph{Submitted}, 2021.

\bibitem{nicholls2022compressive}
D.~Nicholls, A.~Robinson, J.~Wells, A.~Moshtaghpour, M.~Bahri, A.~Kirkland, and N.~Browning, ``Compressive scanning transmission electron microscopy,'' in \emph{Proceedings of the IEEE International Conference on Acoustics, Speech and Signal Processing (ICASSP)}, 2022, pp. 1586--1590.

\bibitem{nicholls2022targeted}
D.~Nicholls, J.~Wells, A.~W. Robinson, A.~Moshtaghpour, M.~Kobylynska, R.~A. Fleck, A.~I. Kirkland, and N.~D. Browning, ``A targeted sampling strategy for compressive cryo focused ion beam scanning electron microscopy,'' \emph{arXiv preprint arXiv:2211.03494}, 2022.

\bibitem{robinson2022sim}
A.~Robinson, D.~Nicholls, J.~Wells, A.~Moshtaghpour, A.~Kirkland, and N.~D. Browning, ``{SIM-STEM} {L}ab: Incorporating compressed sensing theory for fast {STEM} simulation,'' \emph{Ultramicroscopy}, p. 113625, 2022.

\bibitem{robinson2022compressed}
A.~W. Robinson, D.~Nicholls, J.~Wells, A.~Moshtaghpour, A.~I. Kirkland, and N.~D. Browning, ``Compressed {STEM} simulations,'' \emph{Microscopy and Microanalysis}, vol.~28, no.~S1, p. 3116–3117, 2022.

\bibitem{robinson2023towards}
A.~W. Robinson, J.~Wells, D.~Nicholls, A.~Moshtaghpour, M.~Chi, A.~I. Kirkland, and N.~D. Browning, ``Towards real-time {STEM} simulations through targeted subsampling strategies,'' \emph{Journal of microscopy}, vol. 290, no.~1, pp. 53--66, 2023.

\bibitem{stevens2017tensor}
A.~Stevens, Y.~Pu, Y.~Sun, G.~Spell, and L.~Carin, ``Tensor-dictionary learning with deep kruskal-factor analysis,'' in \emph{Artificial Intelligence and Statistics}.\hskip 1em plus 0.5em minus 0.4em\relax PMLR, 2017, pp. 121--129.

\bibitem{zhang2021many}
X.~Zhang, Z.~Chen, and D.~Muller, ``How many detector pixels do we need for super-resolution ptychography?'' \emph{Microscopy and Microanalysis}, vol.~27, no.~S1, pp. 620--622, 2021.

\bibitem{paisley2009nonparametric}
J.~Paisley and L.~Carin, ``Nonparametric factor analysis with beta process priors,'' in \emph{Proceedings of the 26th annual international conference on machine learning}, 2009, pp. 777--784.

\bibitem{zheng2021direct}
Q.~Zheng, T.~Feng, J.~A. Hachtel, R.~Ishikawa, Y.~Cheng, L.~Daemen, J.~Xing, J.~C. Idrobo, J.~Yan, N.~Shibata, Y.~Ikuhara, B.~C. Sales, S.~T. Pantelides, and M.~Chi, ``Direct visualization of anionic electrons in an electride reveals inhomogeneities,'' \emph{Science Advances}, vol.~7, no.~15, p. eabe6819, 2021.

\bibitem{isogai2008formation}
T.~Isogai, H.~Tanaka, T.~Goto, A.~Teramoto, S.~Sugawa, and T.~Ohmi, ``Formation and property of yttrium and yttrium silicide films as low schottcky barrier material for n-type silicon,'' \emph{Japanese journal of applied physics}, vol.~47, no.~4S, p. 3138, 2008.

\bibitem{okunishi2012experimental}
E.~Okunishi, H.~Sawada, and Y.~Kondo, ``Experimental study of annular bright field ({ABF}) imaging using aberration-corrected scanning transmission electron microscopy ({STEM}),'' \emph{Micron}, vol.~43, no.~4, pp. 538--544, 2012.

\bibitem{rodenburg2004phase}
J.~M. Rodenburg and H.~M. Faulkner, ``A phase retrieval algorithm for shifting illumination,'' \emph{Applied physics letters}, vol.~85, no.~20, pp. 4795--4797, 2004.

\bibitem{maiden2009improved}
A.~M. Maiden and J.~M. Rodenburg, ``An improved ptychographical phase retrieval algorithm for diffractive imaging,'' \emph{Ultramicroscopy}, vol. 109, no.~10, pp. 1256--1262, 2009.

\bibitem{maiden2012annealing}
A.~Maiden, M.~Humphry, M.~Sarahan, B.~Kraus, and J.~Rodenburg, ``An annealing algorithm to correct positioning errors in ptychography,'' \emph{Ultramicroscopy}, vol. 120, pp. 64--72, 2012.

\bibitem{maiden2012ptychographic}
A.~M. Maiden, M.~J. Humphry, and J.~M. Rodenburg, ``Ptychographic transmission microscopy in three dimensions using a multi-slice approach,'' \emph{JOSA A}, vol.~29, no.~8, pp. 1606--1614, 2012.

\bibitem{elser2003phase}
V.~Elser, ``Phase retrieval by iterated projections,'' \emph{JOSA A}, vol.~20, no.~1, pp. 40--55, 2003.

\bibitem{d2014deterministic}
A.~D'alfonso, A.~Morgan, A.~Yan, P.~Wang, H.~Sawada, A.~Kirkland, and L.~Allen, ``Deterministic electron ptychography at atomic resolution,'' \emph{Physical Review B}, vol.~89, no.~6, p. 064101, 2014.

\bibitem{bates1989sub}
R.~Bates and J.~Rodenburg, ``Sub-{\aa}ngstr{\"o}m transmission microscopy: a fourier transform algorithm for microdiffraction plane intensity information,'' \emph{Ultramicroscopy}, vol.~31, no.~3, pp. 303--307, 1989.

\bibitem{rodenburg1992theory}
J.~Rodenburg and R.~Bates, ``The theory of super-resolution electron microscopy via {W}igner-distribution deconvolution,'' \emph{Philosophical Transactions of the Royal Society of London. Series A: Physical and Engineering Sciences}, vol. 339, no. 1655, pp. 521--553, 1992.

\bibitem{yang2017electron}
H.~Yang, I.~MacLaren, L.~Jones, G.~T. Martinez, M.~Simson, M.~Huth, H.~Ryll, H.~Soltau, R.~Sagawa, Y.~Kondo, C.~Ophus, P.~Ercius, L.~Jin, A.~Kovács, and P.~D. Nellist, ``Electron ptychographic phase imaging of light elements in crystalline materials using {W}igner distribution deconvolution,'' \emph{Ultramicroscopy}, vol. 180, pp. 173--179, 2017.

\bibitem{martinez2017comparison}
G.~Martinez, M.~Humphry, and P.~Nellist, ``A comparison of phase-retrieval algorithms for focused-probe electron ptychography,'' \emph{Microscopy and Microanalysis}, vol.~23, no.~S1, pp. 476--477, 2017.

\bibitem{o2021contrast}
C.~M. O’Leary, G.~T. Martinez, E.~Liberti, M.~J. Humphry, A.~I. Kirkland, and P.~D. Nellist, ``Contrast transfer and noise considerations in focused-probe electron ptychography,'' \emph{Ultramicroscopy}, vol. 221, p. 113189, 2021.

\bibitem{wang2004image}
Z.~Wang, A.~C. Bovik, H.~R. Sheikh, and E.~P. Simoncelli, ``Image quality assessment: from error visibility to structural similarity,'' \emph{IEEE transactions on image processing}, vol.~13, no.~4, pp. 600--612, 2004.

\bibitem{sertoglu2015scalable}
S.~Sertoglu and J.~Paisley, ``Scalable {B}ayesian nonparametric dictionary learning,'' in \emph{2015 23rd European Signal Processing Conference (EUSIPCO)}, 2015, pp. 2771--2775.

\bibitem{zhou2009non}
M.~Zhou, H.~Chen, J.~W. Paisley, L.~Ren, G.~Sapiro, and L.~Carin, ``Non-parametric {B}ayesian dictionary learning for sparse image representations.'' in \emph{NIPS}, vol.~9, 2009, pp. 2295--2303.

\bibitem{dempster1977maximum}
A.~P. Dempster, N.~M. Laird, and D.~B. Rubin, ``Maximum likelihood from incomplete data via the {EM} algorithm,'' \emph{Journal of the Royal Statistical Society: Series B (Methodological)}, vol.~39, no.~1, pp. 1--22, 1977.

\end{thebibliography}


\begin{thebibliography}{10}
\providecommand{\url}[1]{#1}
\csname url@samestyle\endcsname
\providecommand{\newblock}{\relax}
\providecommand{\bibinfo}[2]{#2}
\providecommand{\BIBentrySTDinterwordspacing}{\spaceskip=0pt\relax}
\providecommand{\BIBentryALTinterwordstretchfactor}{4}
\providecommand{\BIBentryALTinterwordspacing}{\spaceskip=\fontdimen2\font plus
\BIBentryALTinterwordstretchfactor\fontdimen3\font minus \fontdimen4\font\relax}
\providecommand{\BIBforeignlanguage}[2]{{%
\expandafter\ifx\csname l@#1\endcsname\relax
\typeout{** WARNING: IEEEtran.bst: No hyphenation pattern has been}%
\typeout{** loaded for the language `#1'. Using the pattern for}%
\typeout{** the default language instead.}%
\else
\language=\csname l@#1\endcsname
\fi
#2}}
\providecommand{\BIBdecl}{\relax}
\BIBdecl

\bibitem{zheng2021direct}
Q.~Zheng, T.~Feng, J.~A. Hachtel, R.~Ishikawa, Y.~Cheng, L.~Daemen, J.~Xing, J.~C. Idrobo, J.~Yan, N.~Shibata, Y.~Ikuhara, B.~C. Sales, S.~T. Pantelides, and M.~Chi, ``Direct visualization of anionic electrons in an electride reveals inhomogeneities,'' \emph{Science Advances}, vol.~7, no.~15, p. eabe6819, 2021.

\bibitem{sertoglu2015scalable}
S.~Sertoglu and J.~Paisley, ``Scalable {B}ayesian nonparametric dictionary learning,'' in \emph{2015 23rd European Signal Processing Conference (EUSIPCO)}, 2015, pp. 2771--2775.

\bibitem{paisley2009nonparametric}
J.~Paisley and L.~Carin, ``Nonparametric factor analysis with beta process priors,'' in \emph{Proceedings of the 26th annual international conference on machine learning}, 2009, pp. 777--784.

\bibitem{zhou2009non}
M.~Zhou, H.~Chen, J.~W. Paisley, L.~Ren, G.~Sapiro, and L.~Carin, ``Non-parametric {B}ayesian dictionary learning for sparse image representations.'' in \emph{NIPS}, vol.~9, 2009, pp. 2295--2303.

\bibitem{dempster1977maximum}
A.~P. Dempster, N.~M. Laird, and D.~B. Rubin, ``Maximum likelihood from incomplete data via the {EM} algorithm,'' \emph{Journal of the Royal Statistical Society: Series B (Methodological)}, vol.~39, no.~1, pp. 1--22, 1977.

\bibitem{rodenburg2004phase}
J.~M. Rodenburg and H.~M. Faulkner, ``A phase retrieval algorithm for shifting illumination,'' \emph{Applied physics letters}, vol.~85, no.~20, pp. 4795--4797, 2004.

\bibitem{maiden2009improved}
A.~M. Maiden and J.~M. Rodenburg, ``An improved ptychographical phase retrieval algorithm for diffractive imaging,'' \emph{Ultramicroscopy}, vol. 109, no.~10, pp. 1256--1262, 2009.

\bibitem{maiden2012annealing}
A.~Maiden, M.~Humphry, M.~Sarahan, B.~Kraus, and J.~Rodenburg, ``An annealing algorithm to correct positioning errors in ptychography,'' \emph{Ultramicroscopy}, vol. 120, pp. 64--72, 2012.

\bibitem{maiden2012ptychographic}
A.~M. Maiden, M.~J. Humphry, and J.~M. Rodenburg, ``Ptychographic transmission microscopy in three dimensions using a multi-slice approach,'' \emph{JOSA A}, vol.~29, no.~8, pp. 1606--1614, 2012.

\bibitem{elser2003phase}
V.~Elser, ``Phase retrieval by iterated projections,'' \emph{JOSA A}, vol.~20, no.~1, pp. 40--55, 2003.

\bibitem{d2014deterministic}
A.~D'alfonso, A.~Morgan, A.~Yan, P.~Wang, H.~Sawada, A.~Kirkland, and L.~Allen, ``Deterministic electron ptychography at atomic resolution,'' \emph{Physical Review B}, vol.~89, no.~6, p. 064101, 2014.

\bibitem{bates1989sub}
R.~Bates and J.~Rodenburg, ``Sub-{\aa}ngstr{\"o}m transmission microscopy: a fourier transform algorithm for microdiffraction plane intensity information,'' \emph{Ultramicroscopy}, vol.~31, no.~3, pp. 303--307, 1989.

\bibitem{rodenburg1992theory}
J.~Rodenburg and R.~Bates, ``The theory of super-resolution electron microscopy via {W}igner-distribution deconvolution,'' \emph{Philosophical Transactions of the Royal Society of London. Series A: Physical and Engineering Sciences}, vol. 339, no. 1655, pp. 521--553, 1992.

\bibitem{yang2017electron}
H.~Yang, I.~MacLaren, L.~Jones, G.~T. Martinez, M.~Simson, M.~Huth, H.~Ryll, H.~Soltau, R.~Sagawa, Y.~Kondo, C.~Ophus, P.~Ercius, L.~Jin, A.~Kovács, and P.~D. Nellist, ``Electron ptychographic phase imaging of light elements in crystalline materials using {W}igner distribution deconvolution,'' \emph{Ultramicroscopy}, vol. 180, pp. 173--179, 2017.

\bibitem{martinez2017comparison}
G.~Martinez, M.~Humphry, and P.~Nellist, ``A comparison of phase-retrieval algorithms for focused-probe electron ptychography,'' \emph{Microscopy and Microanalysis}, vol.~23, no.~S1, pp. 476--477, 2017.

\bibitem{lozano2018low}
J.~G. Lozano, G.~T. Martinez, L.~Jin, P.~D. Nellist, and P.~G. Bruce, ``Low-dose aberration-free imaging of {L}i-rich cathode materials at various states of charge using electron ptychography,'' \emph{Nano letters}, vol.~18, no.~11, pp. 6850--6855, 2018.

\bibitem{o2021contrast}
C.~M. O’Leary, G.~T. Martinez, E.~Liberti, M.~J. Humphry, A.~I. Kirkland, and P.~D. Nellist, ``Contrast transfer and noise considerations in focused-probe electron ptychography,'' \emph{Ultramicroscopy}, vol. 221, p. 113189, 2021.

\end{thebibliography}
\newpage

\title{Supplemental Material: Simultaneous High-Speed and Low-Dose 4-D STEM Using Compressive Sensing Techniques}
\author{\IEEEauthorblockN{
        Alex W. Robinson\IEEEauthorrefmark{1}\IEEEauthorrefmark{2},
        Amirafshar Moshtaghpour\IEEEauthorrefmark{1}\IEEEauthorrefmark{3},
        Jack Wells\IEEEauthorrefmark{2}\IEEEauthorrefmark{4},
        Daniel Nicholls\IEEEauthorrefmark{1}\IEEEauthorrefmark{2},\\
        Miaofang Chi\IEEEauthorrefmark{5},
        Ian MacLaren\IEEEauthorrefmark{6},
        Angus I. Kirkland\IEEEauthorrefmark{3}\IEEEauthorrefmark{7},
        Nigel D. Browning\IEEEauthorrefmark{1}\IEEEauthorrefmark{2}
    }
    \IEEEauthorblockA{
        \IEEEauthorrefmark{1} Department of Mechanical, Materials and Aerospace Engineering, University of Liverpool, UK.\\
        \IEEEauthorrefmark{2} SenseAI Innovations Ltd., University of Liverpool, Liverpool, UK.\\
        \IEEEauthorrefmark{3} Correlated Imaging Group, Rosalind Franklin Institute, Harwell Science and Innovation Campus, Didcot, UK.\\
        \IEEEauthorrefmark{4}Distributed Algorithms Centre for Doctoral Training, University of Liverpool, UK.\\
        \IEEEauthorrefmark{5}Chemical Science Division, Centre for Nanophase Materials Sciences, Oak Ridge National Laboratory, Tennessee, USA.\\
        \IEEEauthorrefmark{6}School of Physics and Astronomy, University of Glasgow, Glasgow, UK.\\
        \IEEEauthorrefmark{7} Department of Materials, University of Oxford, UK.}
}
\maketitle
\setcounter{section}{0}
\section*{\textbf{\Large{Supplemental Materials}}}
\section{Introduction}
In this supplementary information we provide derivations and mathematical descriptions of the processes which are used in the main text. We start by giving a detailed description of the BPFA reconstruction algorithm, followed by description of our 4-D STEM analysis methods. We also include a section that details the experimental acquisition of the data-set used corresponding to the work of Zheng \etal \cite{zheng2021direct}, as well as a section which details the possible methods for detector sampling, a reconstruction of an interface, and dose distribution maps for estimating dose in STEM.
\section{Beta-Process Factor Analysis (BPFA)}
For the sake of simplified notation, in the remaining of this section we omit the index of the detector pixel $\bs k_{\rm d}$ and identify $\bs Y^{\rm vi}_{\bs k_{\rm d}}$, $\bs X^{\rm vi}_{\bs k_{\rm d}}$, and $\bs N^{\rm vi}_{\bs k_{\rm d}}$ as $\bs Y^{\rm vi}$, $\bs X^{\rm vi}$, and $\bs N^{\rm vi}$, respectively.
Our recovery process adopted from \cite{sertoglu2015scalable} operates as follows for every sub-sampled virtual image in a sequential manner. A sub-sampled virtual image measurements $\bs Y^{\rm vi}$ is first partitioned into $N_{\rm patch}$ overlapping square patches $\{\bs Y^{\rm vi}_{i}\}_{i=1}^{N_{\rm patch}}$ of size $B\times B$ pixels; resulting in $N_{\rm patch} = (H_{\rm p}-B+1)(W_{\rm p}-B+1)$ total number of patches. Similarly, we partition the corresponding virtual image, mask operator, and noise as $\{\bs X^{\rm vi}_{i}\}_{i=1}^{N_{\rm patch}}$, $\{\bs P_{\Omega_{i}}\}_{i=1}^{N_{\rm patch}}$, and $\{\bs N^{\rm vi}_{i}\}_{i=1}^{N_{\rm patch}}$, respectively, such that for each patch $i \in \{1,\cdots,N_{\rm patch}\}$, the sensing model is given by,
\begin{equation} \label{eq:cs-patch-sensing-model}
    \bs Y^{\rm vi}_{i} = \bs P_{\Omega_{i}}( \bs X^{\rm vi}_{i}) + \bs N^{\rm vi}_{i} \in \bb R^{B \times B}.\tag{S1}
\end{equation}

The core of our recovery method assumes that the patches of every virtual image are sparse in a shared dictionary, \ie $\bs x^{\rm vi}_{i} = \bs D \bs \alpha_{i}$, where $\bs x^{\rm vi}_{i} \coloneqq {\rm vec}(\bs X^{\rm vi}_{i}) \in \bb R^{B^2}$ is a vectorised version of $\bs X^{\rm vi}_{i}$, $\bs D \in \bb R^{B^2\times K}$ denotes the dictionary with $K$ atoms and $\bs \alpha_{i} \in \bb R^{K}$ is a sparse vector of weights or coefficients for the $i^{\rm th}$ patch of the virtual image. Based on these definitions, the BPFA algorithm allows us to infer $\bs D$, $\bs \alpha_{i}$, and the noise statistics and in turn reconstruct the virtual images in a sequential fashion.

In summary, the BPFA operates based on the following assumptions. \textit{(i)} The dictionary $\bs D = [\bs d_1^\top,\cdots,\bs d_K^\top]^\top$ has $K$ atoms $\bs d_k \in \bb R^{B^2}$ drawn from a zero-mean multivariate Gaussian distribution. \textit{(ii)} Both the components of the noise vectors $\bs n$ and the non-zero components of the weight vectors $\bs \alpha = \{\bs \alpha_{i}\}_{i=1}^{N_{\rm patch}}$ are drawn \textit{i.i.d.} from zero-mean Gaussian distributions. \textit{(iii)} The sparsity prior on the weights is promoted by the Beta-Bernoulli process\cite{paisley2009nonparametric}. We now let $\bs y^{\rm vi}_i \coloneqq {\rm vec}(\bs Y^{\rm vi}_i) \in \bb R^{B^2}$ be the vectorised version of the sub-sampled virtual image. Hence for all patches $i \in \{1,\cdots,N_{\rm patch}\}$ and dictionary atoms $k \in \{1,\cdots,K\}$, the hierarchy model of BPFA reads
\begin{subequations}\small
\begin{align}
    \bs y^{\rm vi}_i &= \bs P_{\Omega_i} \bs D \bs \alpha_i + \bs n_i, \!& \bs \alpha_i &= \bs z_i \circ \bs w_i \in \bb R^K,\label{eq:bpfa-1}\tag{S2a}\\
    \ts \bs D &= [\bs d_1^\top, \cdots, \bs d_{K}^\top]^{\top}, \!& \bs d_k & \sim  \cl N(0, B^{-2} \bs I_{B^2}),\label{eq:bpfa-2}\tag{S2b}\\
    \bs w_i & \sim  \cl N(0, \gamma_w^{-1} \bs I_{K}), \!& \bs n_i & \sim \cl N(0, \gamma_n^{-1} \bs I_{B^2}),\label{eq:bpfa-3}\tag{S2c}\\
    \bs z_i &\sim \!\ts \prod_{k=1}^{K} {\rm Bernoulli}(\pi_k),\ & \pi_k &\sim\! {\rm Beta}(\ts \frac{a}{K}, \!\ts\frac{b(K-1)}{K}),\label{eq:bpfa-4}\tag{S2d}
\end{align}
\end{subequations}
where $\bs I_K$ is the identity matrix of dimension $K$, $\circ$ denotes the Hadamard product, and $a$ and $b$ are the parameters of the Beta-process. The binary vector $\bs z_i$ in \eqref{eq:bpfa-4} determines the dictionary atoms used to represent $\bs y_i$ or $\bs x_i$; and $\pi_k$ is the probability of using a dictionary atom $\bs d_k$. In \eqref{eq:bpfa-3}, $\gamma_w$ and $\gamma_n$ are the (to-be-inferred) precision or inverse variance parameters. The sparsity level of the weight vectors, $\{\|\bs \alpha_i\|_0\}_{i=1}^{N_{\rm patch}}$ is controlled by the parameters $a$ and $b$ in \eqref{eq:bpfa-4}. However, as discussed in \cite{zhou2009non}, these parameters tend to be non-informative and the sparsity level of the weight vectors is inferred by the data itself.

Unknown parameters in the model above are inferred using stochastic Expectation Maximisation (EM) \cite{dempster1977maximum,sertoglu2015scalable} which involves an expectation step to form an estimation of the latent variables, \ie $\{\bs \alpha_i\}_{i=1}^{N_{\rm patch}}$, and a maximisation step to perform a maximum likelihood estimation to update other parameters. We refer to \cite{sertoglu2015scalable} for a detailed description on the BPFA method for inpainting.
\section{Analysis of 4-D STEM data}
Following acquisition of 4-D STEM data, various techniques such as VDs, DPC, CoM analysis, and phase retrieval techniques such as ptychography can be used for analysis. In all cases, the geometrical centre of the CBED patterns are aligned for consistent analysis. Results of various methods are given in Fig.~S\ref{fig:dpcplot}, Fig.~S\ref{fig:abfplot}, Fig.~S\ref{fig:laadfplot}, Fig. 3 and Fig. 4. 
\subsection{\label{sec:virtualdetectors}Virtual detectors}
A VD is analogous to fixed detectors which are typically used in STEM. A VD, as illustrated in Fig.~1(c), is characterised by inner and outer collection semi-angles $r_{\rm i}$, $r_{\rm o} \in \bb R^{+}$, respectively (in mrad). Given those parameters, we can sum each 2-D CBED patterns over a selected angular range. Setting $\Omega^{\rm vd} \coloneqq \Omega^{\rm vd}(r_{\rm i}, r_{\rm o}) \subset \Omega_{\rm d}$ as the set of detector pixel indices that falls within the radial range of the detector; and letting $\bs Z^{\rm vd} \in \bb R^{H_{\rm p} \times W_{\rm p}}$ be the VD image. Therefore, the value of the VD at probe location $\bs r_{\rm p}$, denoted by $z^{\rm vd}_{\bs r_{\rm p}}$, will be the sum of the 4-D STEM data at probe location $\bs r_{\rm p}$ restricted to the pixels indexed in $\Omega_{\rm vd}$, \ie
\begin{equation}
    z^{\rm vd}_{\bs r_{\rm p}} = \sum_{\bs k_{\rm d} \in \Omega^{\rm vd}} \cl X(\bs r_{\rm p}, \bs k_{\rm d})\tag{S3}\\
\end{equation}


\subsection{\label{sec:dpc}Differential phase contrast}
DPC measures the projected electric field of a sample by quantifying the shift in the electron beam using a segmented (virtual) detector. As depicted in Fig.~1(c), a DPC detector is similar to a VD, but also includes an angular rotation $\theta \in [0,2\pi)$ about the centre of the detector and an angular width $\delta \in [0, 2\pi)$. Let $\Omega^{{\rm dpc}^+} \coloneqq \Omega^{{\rm dpc}^+}(r_{\rm i}, r_{\rm o}, \theta,\delta) \subset \Omega_{\rm d}$ be the set of detector pixel indices whose radii are within the radial range of the detector and whose angles are in the range of $\theta$ and $\theta+\delta$. Similarly, let $\Omega^{{\rm dpc}^-} \coloneqq \Omega^{{\rm dpc}^-}(r_{\rm i}, r_{\rm o}, \theta, \delta)$ be the set of pixel indices whose radii are within the radial range of the detector and whose angles are in the range of $\theta+\pi$ and $\theta+\pi+\delta$. Consequently we define the DPC image by $\bs Z^{\rm dpc} \in \bb R^{H_{\rm p} \times W_{\rm p}}$. Therefore, the value of the DPC image at probe location $\bs r_{\rm p}$, denoted by $z^{\rm dpc}_{\bs r_{\rm p}}$, will be the sum of the CBED pattern at that location and restricted to the pixels indexed in $\Omega^{{\rm dpc}^+}$ minus the sum of that pattern restricted to the pixels indexed in $\Omega^{{\rm dpc}^-}$:
\begin{equation}
    z^{\rm dpc}_{{\bs r}_{\rm p}} = \sum_{\bs k_{\rm d} \in \Omega^{{\rm dpc}^+}} \cl X({\bs r}_{\rm p}, \bs k_{\rm d}) - \sum_{\bs k_{\rm d} \in \Omega^{{\rm dpc}^-}} \cl X(\bs r_{\rm p}, \bs k_{\rm d}).\tag{S4}\\
\end{equation}
\renewcommand{\figurename}{Fig.~S}
\setcounter{figure}{0}

\begin{figure*}[t!]
    \centering
\begin{minipage}{\textwidth}
      \centering
      \begin{minipage}{0.3\linewidth}
\begin{figure}[H]
    \centering
    \scalebox{0.35}{\includegraphics{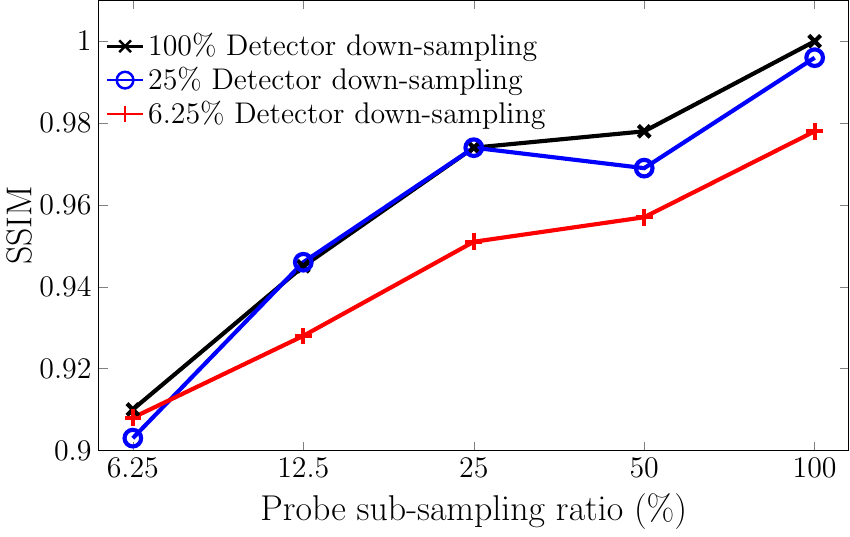}}
    \caption{\textbf{SSIM of DPC images with respect to probe and detector sampling ratios.}}\sq\sq
    \label{fig:dpcplot}
\end{figure}
\end{minipage}
\hspace{0.03\linewidth}
\begin{minipage}{0.3\linewidth}
\begin{figure}[H]
    \centering
    \scalebox{0.35}{\includegraphics{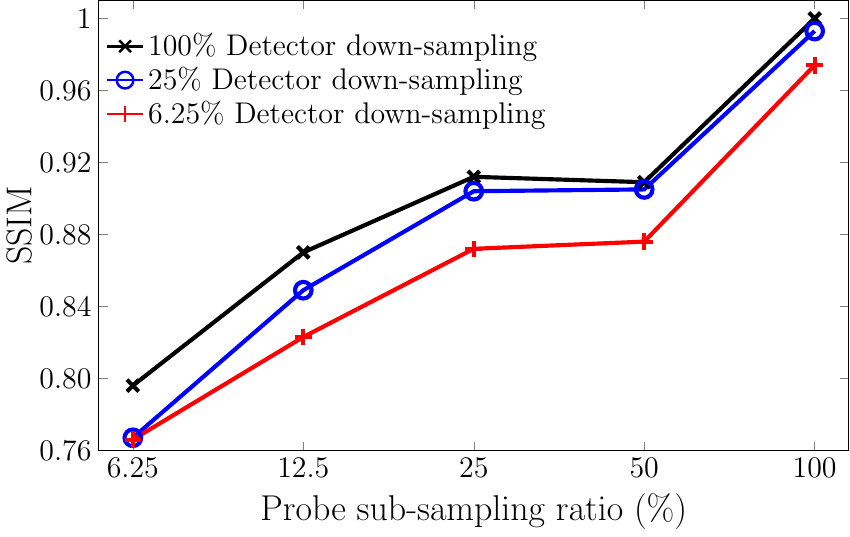}}
    \caption{\textbf{SSIM of ABF images with respect to probe and detector sampling ratios.}}\sq\sq
    \label{fig:abfplot}
\end{figure}
\end{minipage}
\hspace{0.03\linewidth}
\begin{minipage}{0.3\linewidth}
\begin{figure}[H]
    \centering
    \scalebox{0.35}{\includegraphics{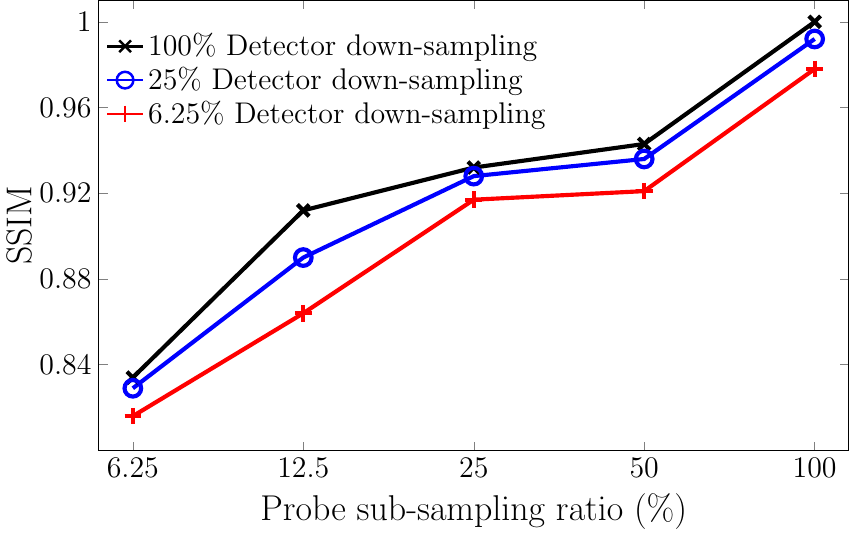}}
    \caption{\textbf{SSIM of LAADF images with respect to probe and detector sampling ratios.}}\sq\sq
    \label{fig:laadfplot}
\end{figure}
\end{minipage}
\end{minipage}
\end{figure*}\label{fig:plots}

\subsection{\label{sec:com}Centre of mass}
The CoM field vector which quantifies the 2-D shift at probe location $\bs r_{\rm p}$ is denoted by $\bs z^{\rm com}_{\bs r_{\rm p}} \in \bb R^{2}$ to construct a full CoM vector field $\bs Z^{\rm com} \in \bb R^{H_{\rm p}\times W_{\rm p} \times 2}$.
Let $\Omega^{\rm bd} \coloneqq \Omega^{\rm bd}(\bs k_{\rm d}) \subset \Omega_{\rm d}$ be the set of detector pixel indices that falls within the bright field disk. We assume that each CBED pattern can be modelled as a non-uniform density lamina where the density is equivalent to the intensity of the signal in the CBED pattern. We then use standard derivations to derive the CoM field coefficients, $\bs z^{\rm com}_{\bs r_{\rm p}}$, as
\begin{equation}
    \bs z^{\rm com}_{\bs r_{\rm p}} = \frac{\sum_{\bs k_{\rm d} \in \Omega^{\rm bd}} (\bs k_{\rm d} - \bs c_{\rm d})\cdot \cl X(\bs r_{\rm p}, \bs k_{\rm d})}
    {\sum_{\bs k_{\rm d} \in \Omega^{\rm bd}} \cl X(\bs r_{\rm p}, \bs k_{\rm d})},\tag{S5}\\
\end{equation}
where $\bs c_{\rm d} \in \bb R^{2}$ are the coordinates of the centre of the CBED pattern. Following this, the CoM displacement can be given as the magnitude or the angle of the vector $\bs z^{\rm com}_{\bs r_{\rm p}}$.Furthermore, the projected electric charge density is given as the divergence of the CoM displacement field (including a field and charge constant terms).
\subsection{\label{sec:ptycho}STEM ptychography}
Ptychography is a technique that recovers the complex object wavefunction illuminated by a (partially) coherent source, which in the case of STEM is a focused or intentionally defocused probe. There are a number of analytical and iterative algorithms\cite{rodenburg2004phase,maiden2009improved,maiden2012annealing,maiden2012ptychographic,elser2003phase,d2014deterministic} that recover the wavefunction and in this work we  use an adaptation of the Wigner distribution deconvolution (WDD)~\cite{bates1989sub,rodenburg1992theory} which is one method for object phase recovery for focused probe illumination~\cite{yang2017electron,martinez2017comparison,lozano2018low,o2021contrast}. 

We firstly introduce a definition of the observed CBED patterns as,
\begin{equation}
    \cl X(\bs{r}_{p},\bs{k}_{d}) = \lvert \cl I(\bs{r}_{p},\bs{k}_{d})\rvert^{2}\tag{S6}\\
\end{equation}
where,
\begin{equation}
    \cl I(\bs{r}_{p},\bs{k}_{d}) = \int P(\bs{r} - \bs{r}_{p})o(\bs{r})\exp{(i2\pi \bs{r}\cdot \bs{k}_{d})}d\bs{r}\tag{S7}\\
\end{equation}
which implies that $\cl X$ is a convolution between the object transfer function $o(\bs{r})$ and probe function $P(\bs {r})$. To recover the object phase, we calculate the $\cl H$-matrix (or $\cl H$-array, for the sake of consistent notation), which is the Fourier transform of a 4-D STEM data-set with respect to real space probe locations, followed by an inverse Fourier transform with respect to the detector pixel locations, \ie 

\begin{equation} \label{eq:h_array_definition}
    \cl H(\bs k_{\rm p},\bs r_{\rm d}) = \cl F^{-1}_{\bs k_{\rm d}}\Big[\cl F_{\bs r_{\rm p}}\big[\cl X(\bs r_{\rm p}, \bs k_{\rm d})\big]\Big],\tag{S8}\\
\end{equation}
where $\bs k_{\rm p}$ are the reciprocal space coordinates of the probe locations and $\bs r_{\rm d}$ are real space coordinates with respect to the detector pixels.
\begin{figure*}[t!]
    \centering
    \scalebox{0.55}{\includegraphics{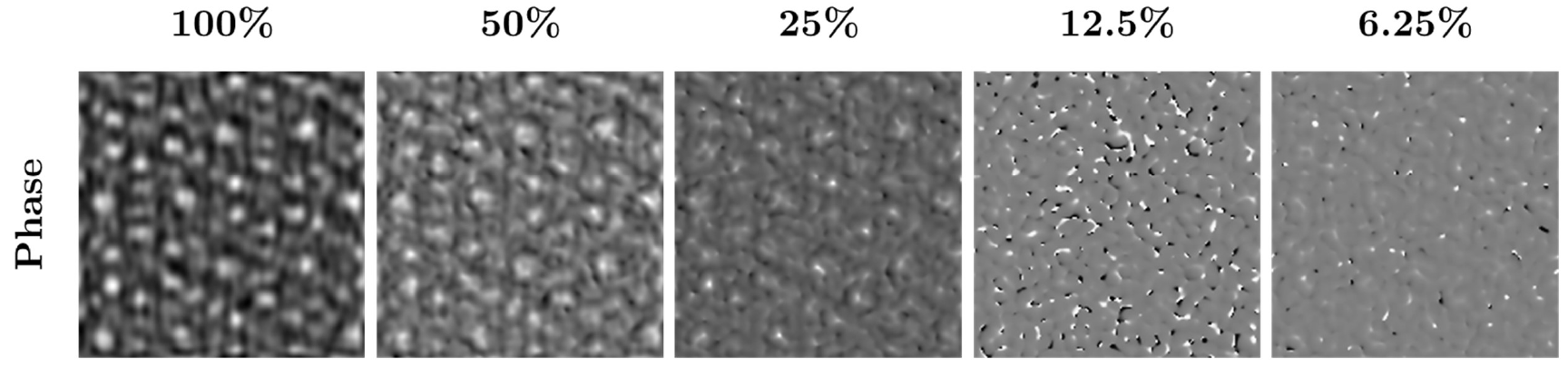}}
    \caption{\textbf{WDD object phase applied to probe sub-sampled data (indicated by column heading) without inpainting.}}\sq\sq
    \label{fig:noinpaint}
\end{figure*}
For a general function $f(\bs u)$, we define its Wigner distribution\cite{bates1989sub,rodenburg1992theory} as,
\begin{equation}\label{eq:wigner_distribution}
\cl W_{f}(\bs u, \bs v) = \cl F^{-1}_{\bs v'} \big[f(\bs u + \bs v') \cdot f^*(\bs v')\big].\tag{S9}\\
\end{equation}
Using this definition of a Wigner distribution function in \ref{eq:wigner_distribution}, it can be shown that the $\cl H$-array is the product of two Wigner distributions corresponding to the probe $\cl W_{p}$ and object $\cl W_{o}$, \ie
\begin{equation}
    \cl H(\bs k_{\rm p},\bs r_{\rm d}) = \cl W_{P}(-\bs k_{\rm p},\bs r_{\rm d})\cdot \cl W_{O}(\bs k_{\rm p},\bs r_{\rm d}),\tag{S10}\\
\end{equation}
where $\cl W_{P}(\bs k_{\rm p}, \bs r_{\rm d})$ is estimated as the initial probe parameters.

The Wigner distribution of the object transfer function in the reciprocal space can then be computed by a Wiener deconvolution routine, with the inclusion of a small constant $\epsilon >0$ to avoid division by zero, as
\begin{equation}\label{eq:wigner_object}
    \cl W_{O}(\bs k_{\rm p},\bs r_{\rm d}) = \frac{\cl W_{P}^{*}(-\bs k_{\rm p},\bs r_{\rm d})\cl H(\bs k_{\rm p},\bs r_{\rm d})}{\lvert \cl W_{P}(-\bs k_{\rm p},\bs r_{\rm d})\rvert^{2} + \epsilon} .\tag{S11}\\
\end{equation}
Once $\cl W_{O}(\bs k_{\rm p},\bs r_{\rm d})$ is computed  in \eqref{eq:wigner_object}, we can write
\begin{equation}\label{eq:L-matrix}
    O^*(\bs k_{\rm d}) \cdot O(\bs k_{\rm p} + \bs k_{\rm d}) = \cl L(\bs k_{\rm p}, \bs k_{\rm d}) \coloneqq \cl F_{\bs r_{\rm d}} \big[\cl W_{O}(\bs k_{\rm p},\bs r_{\rm d})\big],\tag{S12}\\
\end{equation}
where $O(\bs k_{\rm p}) = \cl F_{\bs r_{\rm p}}\big[o(\bs r_{\rm p})\big]$ is the Fourier transform of the object transfer function as a function of the spatial frequency of the probe location. It is clear from \eqref{eq:L-matrix} that $| O(\bs 0)|^2  = \cl L(\bs 0, \bs 0)$; and therefore,
\begin{equation}\label{eq:object_wdd_fourier}
    O(\bs k_{\rm p}) = \frac{\cl L(\bs k_{\rm p},\bs 0)}{\sqrt{\cl L(\bs 0, \bs 0)} e^{j\theta_0}},\tag{S13}\\
\end{equation}
where $\theta_0$ is the phase of the Fourier transform of the object transfer function at $\bs k_{\rm p} = \bs 0$. Finally, an inverse Fourier transform on $O(\bs k_{\rm p})$ yields the object transfer function in the probe location coordinates:
\begin{equation}\label{eq:object_wdd}
    o(\bs r_{\rm p}) = \cl F^{-1}_{\bs k_{\rm p}}\big[O(\bs k_{\rm p})\big].\tag{S14}\\
\end{equation}
We note that the term $e^{j\theta_0}$ in \eqref{eq:object_wdd_fourier} causes a global relative phase shift in the estimation of the Fourier transform of the object transfer function, equivalent to a spatial shift in real space. Without loss of generality, we can set $\theta_0 = 0$. We also note that the estimated object transfer function recovered using the WDD in \eqref{eq:object_wdd} is a function of $\bs r_{\rm p}$, \ie the real space coordinates of the probe location. Therefore, regardless of the number of detector pixels, the WDD estimation of the object transfer function has the same dimensionality as the scanning grid.

\begin{figure*}[t!]
    \centering
    \scalebox{0.7}{\includegraphics{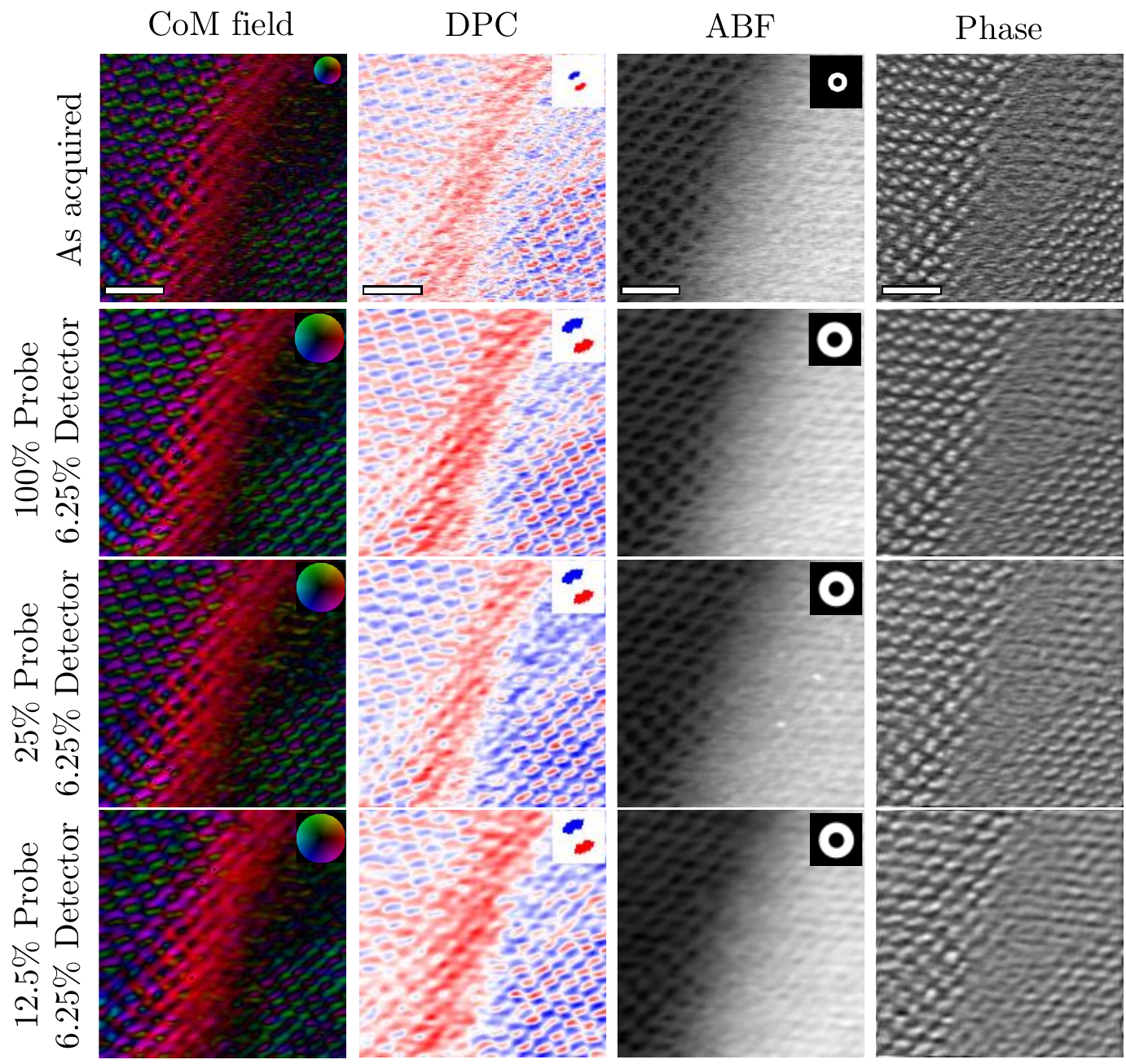}}
    \caption{\textbf{Simulation of sub-sampled 4-D STEM using experimentally acquired 4-D STEM data of a CdTe-Si interface.} Top row shows the ABF, DPC, CoM field, and object phase reconstruction using WDD (from left to right) for the fully sampled, raw data. The remaining rows are then down-sampled on the detector (6.25\%) and probe subsampled, with the recovery of the data being performed using the BPFA. Scale bar indicates 1nm.}\sq\sq
    \label{fig:fig_cdte_4d_results}
\end{figure*}

\subsection{Practical acquisition of yttrium silicide data}

The 4-D STEM data was acquired using an aberration-corrected Nion UltraSTEM100 equipped with a cold field-emission electron source, operated at an accelerating voltage of $100$ kV. The probe is $\sim20$ pA with a convergence semi-angle of $30$ mrad. In this data, the Hamamatsu ORCA ultra-low noise scientific CMOS sensor with a $2048\times2048$ readout display was chosen to readout $256\times256$ pixels then binned to $128\times128$ in software. This gave a readout speed of $760$fps. Further information can be found in \cite{zheng2021direct}.

\subsection{Detector sampling}

In most cases, detector read-out speed is limited by the minimal read-out region of the detector. A global shutter can be selected where the total frame is acquired over the same time period, however the data can only be readout and stored at a rate proportional to the size of the readout region. 

In order to get around this, users can reduce the camera length such that the CBED pattern is contained within the readout region, or an idealised detector would allow for an arbitrary readout region. Furthermore, newer event based detectors would allow for faster readout. 

\subsection{Acquisition speed}

Consider an arbitrary detector with a fixed maximum readout frequency of $f$Hz. Next, assume a scanning grid of size $H_{\rm{p}}\times W_{\rm{p}}$, such that the time to perform the full raster scan, $T\in\bb R_{+}$, to collect diffraction patterns is:

\begin{equation}
T = \frac{H_{\rm{p}}W_{\rm{p}}}{f}~~.
\end{equation}

Now, assume that only a subset of diffraction patterns are collected according to a sampling operator, $\Omega$, where the number of acquired diffraction patterns over the same field of view is $M<H_{\rm{p}}W_{\rm{p}}\in\bb N$. This then implies that the time to acquire the subset, $\hat{T}\in\bb R_{+}$, of diffraction patterns is then:

\begin{equation}
\hat{T} = \frac{M}{f}~~,
\end{equation}
which by virtue of $M<H_{\rm{p}}W_{\rm{p}}$, implies that $\hat{T}<T$.  

This differs to the work reported by Stevens \etal~(2018) by virtue of the recovery method, indicating a lower sampling ratio of the data, a different strategy for detector sampling and a generalisation to any counting or integrating detector.

\begin{figure*}[t!]
    \centering
    \scalebox{0.45}{\includegraphics{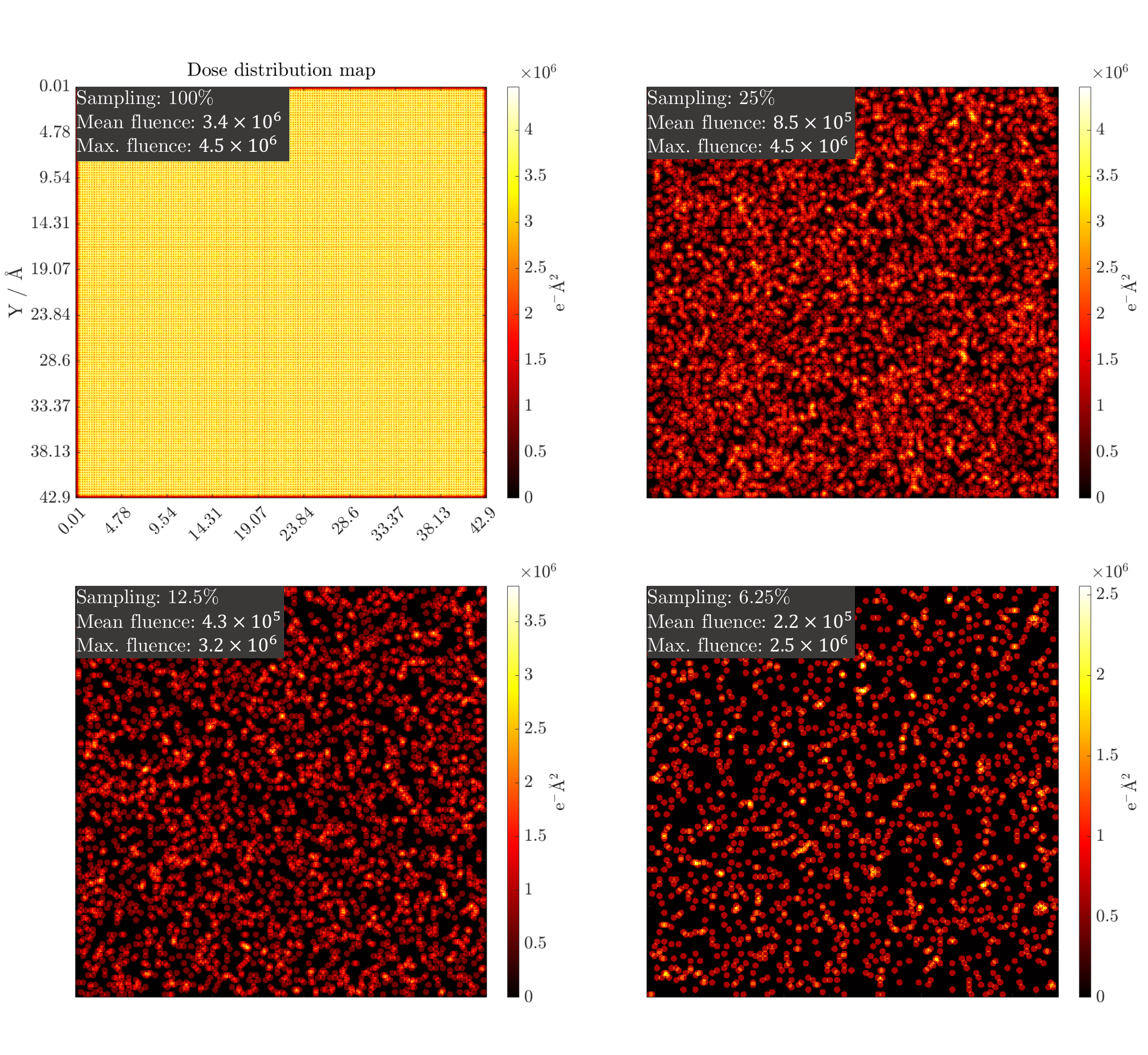}}
    \caption{Dose distribution maps for UDS sampling in a STEM with the parameters according to those used in Fig.~S\ref{fig:fig_cdte_4d_results}. The results show that the average fluence drops according to sampling rate, and the maximum fluence also drops beyond a certain sampling rate. The probe is assumed to have a radius of 0.035nm.}\sq\sq
    \label{fig:fig_dose_distribution}
\end{figure*}

\subsection{Method applied to a heterogeneous interface}

A test data set of a cadmium-telluride/silicon interface was acquired using a JEOL JEM-ARM200F at the Institute for Microelectronics and Microsystems in Catania. The microscope was aligned with a focused STEM probe at 200kV and a convergence semi-angle of 30 mrad. The scan-step was set to 0.025nm, and the detector pixel size was calculated as 1 mrad, with a maximum collection semi-angle of 64 mrad. Diffraction patterns of size 128$\times$128 (binned in software from $1024\times1024$) were collected over a raster scan containing 170$\times$170 probe positions. To recover the phase image, the Wigner Distribution Deconvolution (WDD) algorithm was chosen. 

Results of Fig.~S\ref{fig:fig_cdte_4d_results} show that the same method can be applied to complex, aperiodic structures, in this case to a 12.5\% sampling ratio of probe locations, and 6.25\% in detector sampling. This number is not as low as for the data in the main document, however the scan step here is significantly higher.

The probe current was approximately 14pA, and the dwell time was limited by the camera readout speed to 2.5ms. This gives an estimated electron fluence of $3.4\times10^{8}$e$^{-}$nm$^{-2}$. Electron fluence distribution maps are given in Fig.~S\ref{fig:fig_dose_distribution} for various sampling ratios, but the same dwell time and probe current.

\subsection{BPFA applied to a defect rich structure}

To further demonstrate the BPFA algorithm's robustness to complex structures, an image is constructed containing various types of defects. These defects include an interstitial dopant, a vacancy, a screw dislocation, a lattice distortion, and a grain boundary. Furthermore, each of the `atoms' have randomly assigned intensity to further complicate the structure. In the BPFA algorithm, each overlapping patch of the image is inpainted independently. This means that periodic structures are equivalent to aperiodic structures in terms of how the algorithm performs the inpainting. Each of the overlapping patches is unaware of the global structure, and the only connection between the overlapping patches is through the shared dictionary. Assuming parameters are carefully chosen, the patches will not be inpainted with artefacts caused by defects in other parts of the image. It is for this exact reason why the algorithm is chosen, since it is able to then inpaint local variations (\ie defects) without requiring a training data of purely defects. The algorithm will combine dictionary elements to minimise the residual for that overlapping patch, and as long as the correct parameters are chosen, the recovery is robust to defects.

\begin{figure*}[t!]
    \centering
    \scalebox{0.62}{\includegraphics{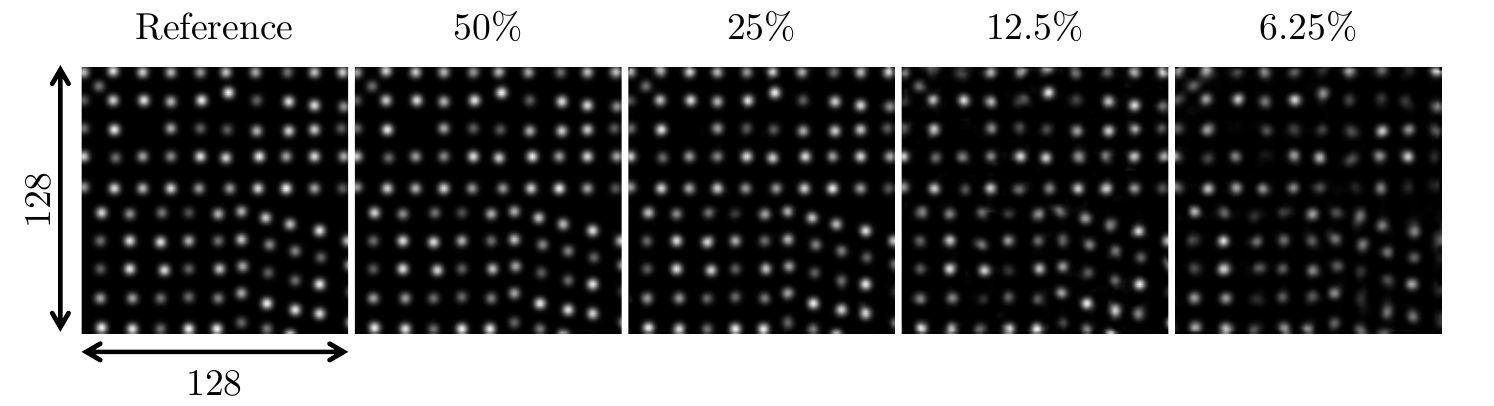}}
    \caption{\textbf{Testing the BPFA algorithm on a complex structure.} Inpainting results at various sampling ratios (above each reconstruction) for the complex structure containing various defects such as an interstitial dopant, a vacancy, a lattice distortion, a grain boundary, and a screw dislocation. The radii of the atoms in the structure are approximately 3 - 3.5 pixels, which is equivalent to roughly a 0.025nm~-0.035nm~scan step.}\sq\sq
    \label{fig:fig_complex_structure}
\end{figure*}

\end{document}